\documentclass[fleqn,usenatbib]{mnras}

\usepackage{newtxtext,newtxmath}
\usepackage[T1]{fontenc}

\DeclareRobustCommand{\VAN}[3]{#2}
\let\VANthebibliography\thebibliography
\def\thebibliography{\DeclareRobustCommand{\VAN}[3]{##3}\VANthebibliography}

\usepackage{graphicx}	% Including figure files
\usepackage{amsmath}	% Advanced maths commands
\usepackage{xcolor}     % For highlighting changes

\usepackage{verbatim}

\title[Parametric instability feedback on warped discs]{Parametric instability in warped astrophysical discs: growth, saturation and feedback}

\author[Fairbairn \& Ogilvie]{
Callum W. Fairbairn$^{1}$\thanks{E-mail: cwf29@cam.ac.uk}
and Gordon I. Ogilvie$^{1}$\thanks{E-mail: gio10@cam.ac.uk}
\\
$^{1}$Department of Applied Mathematics and Theoretical Physics, University of Cambridge, Centre for Mathematical Sciences,\\
Wilberforce Road, Cambridge CB3 0WA, UK
}

\date{Submitted September 2022}

\pubyear{2022}

\begin{document}
\label{firstpage}
\pagerange{\pageref{firstpage}--\pageref{lastpage}}
\maketitle

\begin{abstract}
Attempts to understand the dynamics of warped astrophysical discs have garnered significant attention, largely motivated by the growing catalogue of observed distorted systems. Previous studies have shown that the evolution of the warp is crucially regulated by the internal flow fields established by the undulating geometry. These are typically modelled as laminar horizontal, shearing flows which oscillate back and forth at approximately the orbital frequency. However this shearing motion is known to be susceptible to a hydrodynamic, parametric instability of inertial waves which might modify the warped dynamics. Whilst the linear growth phase is well understood, the subsequent nonlinear saturation combined with the self-consistent feedback onto the warp has not been studied. In this work, we implement a novel numerical setup using the recent ring model framework of Fairbairn and Ogilvie, within the Lagrangian code \texttt{GIZMO}. We formally identify several locally growing modes in the simulation, as predicted by a three-mode coupling analysis of the instability, and find decent agreement with the theoretical growth rates. We understand the saturation mechanism as a wave breaking process which suppresses the growth of shorter wavelength parametric couplings first, whilst allowing the longest mode to dominate the final quasi-steady, wavelike turbulence. The Reynolds stresses, transporting energy from the warp to the small scales, can be effectively modelled using a time-dependent, anisotropic viscous alpha model which closely captures the amplitude and phase evolution of the warp. Consequently, this model might help inform future global studies which are commonplace but typically don't resolve the parametric instability.
\end{abstract}


%%%%%%%%%%%%%%%%%%%%%%%%%%%%%%%%%%%%%%%%%%%%%%%%%%

% %TC:ignore
% \section*{Word Counts}

% This section is \textit{not} included in the word count.

% \quickwordcount{main}

% \quickcharcount{main}

% \detailtexcount{main}
% %TC:endignore

%%%%%%%%%%%%%%%%% BODY OF PAPER %%%%%%%%%%%%%%%%%%

%===========================================%
\section{Introduction}
\label{section:introduction}
%===========================================%

%-------------------------------------------%
\subsection{Astrophysical motivation}
\label{subsection:astrophysical_motivation}
%-------------------------------------------%

Warped discs are ubiquitous in a range of astrophysical contexts and arise whenever there is a misalignment present in the system. This could be due to the randomised accretion of material onto the nascent disc around a young star from a turbulent molecular cloud \citep{LucasEtAl2013,Bate2018}. Otherwise, a tilted magnetic field from the central star can interact with the disc and excite inclinations \citep{Lai1999} or the Lense Thirring torque from a spinning black hole can establish an undulating disc profile \citep{BardeenPetterson1975}. Furthermore, planetary or binary stellar companions on inclined orbits can gravitationally torque the disc and lead to disc warps and even more extreme phenomena like disc tearing and breaking \citep{NixonKing2012} as seen in a variety of simulations \citep[e.g.][]{FacchiniEtAl2013, NealonEtAl2016, Zhu2019}.

Such warps were originally inferred indirectly by the super-orbital periods of X-ray binaries such as Hercules X-1 \citep{Katz1973}. As the distorted disc precesses in front of the line of sight, the signal is modulated \citep[e.g.][]{KotzeCharles2012,PoonEtAl2021}. A similar effect has been seen in protoplanetary discs wherein the central misaligned regions attenuate light from the stellar source and cast shadows onto the outer disc \citep[e.g.][]{DebesEtAl2017,Muro-ArenaEtAl2020}. Recently, direct observations have also been made wherein the midplane of IRAS 04368+2557 tracks a modest warp \citep{SakaiEtAl2019}. Finally, the triple star system of GW Orionis presents compelling evidence for multiple broken and tilted rings which have been torn apart through the complex gravitational interactions \citep{KrausEtAl2020}. With the current observational revolution exploiting the complementary ground and spaced based missions of the \textit{Atacama Large Millimeter/submillimeter Array} (ALMA) and the \textit{James Webb Space Telescope} (JWST), we expect this catalogue of warped systems to continue to grow. 

%-------------------------------------------%
\subsection{Theoretical background}
\label{subsection:theoretical_background}
%-------------------------------------------%

Of course, this phenomenology demands an accompanying theoretical understanding. The foundations of warp dynamics were proposed by Petterson who modelled the warp as a series of nested, viscously interacting rings \citep{Petterson1977a,Petterson1977b,Petterson1978}. This led to a diffusion and damping of the warp on a viscous time-scale. However \cite{PapaloizouPringle1983} pointed out the significance of internal oscillatory shear flows driven by pressure gradients which result from the warped geometry. These efficiently advect angular momentum such that the warp diffuses and damps more rapidly when $H/R < \alpha < 1$, where $\alpha$ is the Shakura-Sunyaev viscosity parameter \citep{ShakuraSunyaev1973} and $H/R$ is the angular semi-thickness of the disc. Subsequently, \cite{PapaloizouLin1995} showed that linear warps propagate as bending waves when $\alpha < H/R$. Attempts to extend this understanding into the nonlinear regime were pioneered by \cite{Pringle1992} before \cite{Ogilvie1999} developed a fully nonlinear, self-consistent model of diffusion in Keplerian discs and bending waves in non-Keplerian discs. Furthermore, efforts to understand the resonant Keplerian, inviscid regime have been explored by \cite{Ogilvie2006} who developed evolutionary equations which follow a propagating, weakly nonlinear bending wave. More recently, \cite{FairbairnOgilvie2021a} have proposed a novel local ring model which also predicts the fully nonlinear extension of precessing bending modes \citep{FairbairnOgilvie2021b}. 

Despite all these efforts, the theories typically simplify matters by assuming laminar internal flows. Indeed, any small scale turbulence is mysteriously encapsulated by the viscous $\alpha$ parameter, as it is in many numerical simulations in which the flow remains laminar. This $\alpha$ is thought to owe its existence to some underlying instability in the disc but its magnitude and self-consistent dependence on disc properties are still poorly understood. One such hydrodynamic mechanism thought to be active in warped discs is the parametric instability. This feeds off the free energy contained within the oscillating shear flows, facilitated by a mode coupling resonance between the shear flow associated with the warp and a pair of inertial waves \citep{PapaloizouTerquem1995,GammieEtAl2000}. This has been verified in the local warped shearing box analysis of \cite{OgilvieLatter2013b} wherein an imposed warp, built into the coordinate system, drives the oscillating shear flow. Subsequently, this model was numerically implemented by \cite{PaardekooperOgilvie2019a} who obtained the linear growth rates and found a nonlinear saturation for viscous discs with a fixed warp. However, a notable limitation of this warped shearing box framework is that the amplitude of the linear shear flow is tempered only by the viscosity, since no time-dependence of the warp is allowed for. A similar local model has been developed by \cite{OgilvieBarker2014} to analyse the growth of the parametric instability in the parallel problem of eccentric, distorted discs \citep{BarkerOgilvie2014,Papaloizou2005a,Papaloizou2005b}. Subsequent 2D nonlinear simulations in this eccentric shearing box were performed by \cite{WienkersOgilvie2018}, where they tried to understand the saturation of the inertial modes as a result of wave breaking.

Despite these local efforts most global simulations fail to observe the parametric instability in warped systems. Indeed the computational challenges associated with an undulating geometry mean that the simulation is unavoidably three dimensional with no natural grid geometry and the required vertical extent of the domain leads to very low density regions. Meanwhile, eccentric discs remain coplanar and grid based codes can be used to greater effect. Indeed, global simulations of \citet{PierensEtAl2020} found that the parametric instability is triggered in an eccentric circumbinary disc and affects the conditions for planet formation. In contrast, Lagrangian based codes such as smoothed particle hydrodynamics (SPH) \citep{GingoldMonaghan1977,Lucy1977} are most commonly employed to study warped disc dynamics. Indeed, SPH studies have been used to great effect, demonstrating excellent agreement with the linear theory of warp propagation and diffusion \citep[e.g.][]{LodatoPrice2010,FacchiniEtAl2013} as well as probing the nonlinear effects of disc tearing and breaking \citep[e.g.][]{LarwoodPapaloizou1997,NixonKing2012,RajEtAl2021}. 

However, the glaring lack of the parametric instability in such global warped simulations is worrying. This might owe to a lack of resolution below the length scale of the instability, as discussed in \cite{PaardekooperOgilvie2019a}. Alternatively, it might be suggested that the parametric instability cannot exist in global evolving warps wherein the coherency of local boxes with periodic boundary conditions is removed. However, in a recent breakthrough simulation, \cite{DengEtAl2020} used an unprecedented 120 million particles in a low viscosity, Lagrangian Godunov scheme \citep{Hopkins2015}. They finally observed the emergence of the parametric instability which was found to significantly damp the warp within a few bending wave crossing times. This underlines the importance of the parametric instability and necessitates a means to better incorporate its effect in future analytical and global numerical efforts. Indeed, it is unclear whether the simple $\alpha$ prescription is qualitatively well suited to describe the parametric instability and there are only tentative estimates as to its quantitative magnitude in the ensuing non-linearly saturated state \citep{PaardekooperOgilvie2019a}.

%-------------------------------------------%
\subsection{Outline of this paper}
\label{subsection:outline}
%-------------------------------------------%

In this work we aim to provide the first detailed local simulations of the growth and nonlinear saturation of the parametric instability in a freely evolving warped disc. This facilitates the slow time dependence of the precessing warp in the inertial frame, despite the disc being Keplerian. Furthermore, by allowing for the feedback onto the warp, we crucially capture a self-consistent picture of the damping process which causes the warp amplitude to evolve. We will compare this behaviour with some simple viscous models to explore how the dynamics might be incorporated in future analytical or globally unresolved numerical studies. We begin in Section \ref{section:ring_model} by summarising the local framework which we use to model a freely evolving Keplerian warped disc. The numerical implementation is then described and tested in Section \ref{section:numerical}. The emergence and growth of the parametric instability is analysed in Section \ref{section:instability} before we model the resulting feedback onto the warp dynamics in Section \ref{section:feedback}. Finally, we will discuss our findings in Section \ref{section:discussion} and present our conclusions in Section \ref{section:conclusion}.  
%===========================================%
\section{Summary of ring model framework}
\label{section:ring_model}
%===========================================%

In this work we motivate our numerical setup, based on the ring model developed in \cite{FairbairnOgilvie2021a} and \cite{FairbairnOgilvie2021b}, hereafter FOA and FOB respectively. This framework proved useful when investigating nonlinear solutions for warped disc dynamics and allows one to consider the self-consistent evolution of warp in the troublesome Keplerian regime where resonances complicate the dynamics. Here we will briefly summarise the model but refer the reader to FOA for a detailed explanation.

We switch into a shearing-box reference frame and expand the Navier-Stokes equations about some reference orbit assuming an axisymmetric potential $\Phi(r,z)$. Indeed, \cite{Ogilvie2022} provides a thorough justification for the use of 2D local models in studying warped disc dynamics. This procedure neglects curvature effects so the geometry is locally Cartesian, with the radial, azimuthal and vertical directions denoted $(x,y,z)$ respectively. The momentum equation becomes
\begin{equation}
\label{eq:local_euler_equation}
    \frac{D \mathbf{u}}{Dt} +2\bmath{\Omega}\times\bmath{u} = -\nabla(\Phi_{\mathrm{t}}+h)+T\nabla s,
\end{equation}
whilst the thermodynamic equations are
\begin{equation}
\label{eq:local_thermal_equation}
    \frac{D h}{D t}  = -(\gamma-1)h \nabla\cdot\bmath{u} \quad \text{and} \quad \frac{D s}{D t} = 0,
\end{equation}
where the material derivative $D/Dt \equiv \partial_t + \mathbf{u}\cdot\nabla$. Here $\mathbf{u}$ denotes the velocity vector, $\mathbf{\Omega} = \Omega \hat{\mathbf{z}}$ is the orbital frequency,  $\gamma$ is the adiabatic index, $T$ is the temperature, $s$ is the specific entropy and $h$ is the specific enthalpy
\begin{equation}
    \label{eq:enthalpy}
    h = \gamma e = \frac{\gamma p}{(\gamma-1)\rho}.
\end{equation}
This is simply proportional to the specific internal energy $e$ which is given in terms of the ratio of pressure $p$ to density $\rho$. The local expansion of the tidal potential is given by 
\begin{equation}
    \label{eq:tidal_potential}
    \Phi_{\mathrm{t}} = -\Omega S x^2+\frac{1}{2} \nu^2 z^2, 
\end{equation}
where $S = -(r \textrm{d}\Omega/\textrm{d}r)$ is the orbital shear rate whilst $\nu^2$ is the squared vertical oscillation frequency of a test particle perturbed from a circular orbit.  

To make analytical progress, we simplify matters by assuming axisymmetric solutions for which the flow velocity is laminar and linear in the meridional coordinates, i.e. we adopt $u_i = A_{ij}x_j$ where the subscript indices $(1,2,3)$ correspond to the $(x,y,z)$ coordinate directions and $A_{ij}$ is the so called flow matrix. Meanwhile, the pressure/density structure is posited to be some quadratic function such that the cross-sectional isobars are elliptical and centred on the origin. With these simplifications, the system can be reduced to a simple set of ordinary differential equations. Here we will motivate these from a Lagrangian perspective which naturally lends itself to our future numerical experiments which track individual particles. Consider a reference state $\mathbf{x_0} = (x_0,y_0,z_0)$ in which mass is distributed such that the materially conserved density and pressure contours lie on circles with radius $L R$. Here, $R$ is a dimensionless radial coordinate measured in units of the characteristic length $L$, which is defined as the second moment of the reference mass distribution. Then the density and pressure in the reference state have the form $\rho_0 = \hat{\rho}_0 \tilde{\rho}(R)$ and $p_0 = \hat{p}_0 \tilde{p}(R)$ respectively, where $\tilde{\rho}$ and $\tilde{p}$ are suitably normalized dimensionless functions and $\hat{\rho}_0$ and $\hat{p}_0$ are dimensional constants. This stationary reference configuration is then mapped to the elliptical, dynamical state $\mathbf{x}$ via a time dependent linear transformation $\mathbf{x} = \mathbf{J}\mathbf{x_0}$, as encapsulated within the $J_{ij}$ matrix. The ring evolution is then captured by the evolution of these 6 components:
\begin{align}
    \label{eq:J11_ode}
    & \ddot{J}_{11} = 2\Omega\dot{J}_{21}+2\Omega S J_{11}+\frac{\hat{T}_0}{J^\gamma L^2}J_{33}, \\
    \label{eq:J13_ode}
    & \ddot{J}_{13} = 2\Omega\dot{J}_{23}+2\Omega S J_{13}-\frac{\hat{T}_0}{J^\gamma L^2}J_{31}, \\
    \label{eq:J21_ode}
    & \ddot{J}_{21} = -2\Omega\dot{J}_{11}, \\
    \label{eq:J23_ode}
    & \ddot{J}_{23} = -2\Omega\dot{J}_{13}, \\
    \label{eq:J31_ode}
    & \ddot{J}_{31} = -\nu^{2}J_{31}-\frac{\hat{T}_0}{J^\gamma L^2}J_{13}, \\
    \label{eq:J33_ode}
    & \ddot{J}_{33} = -\nu^{2}J_{33}+\frac{\hat{T}_0}{J^\gamma L^2}J_{11}.
\end{align}
Here, $J =\det(J_{ij})=J_{11}J_{33}-J_{13}J_{31}$ is the Jacobian determinant and is proportional to the cross-sectional ellipse area and $\hat{T}_0 = \hat{p}_0/\hat{\rho}_0$ is a characteristic temperature governing the pressure support of the ring. We can exploit the integrability of equations \eqref{eq:J13_ode} and \eqref{eq:J23_ode}, which is a consequence of the conservation of angular momentum, to eliminate $J_{21}$ and $J_{23}$ yielding
\begin{align}
   & \ddot{J}_{11} +\kappa^2 J_{11}=2C_z+\frac{\hat{T}_0}{J^\gamma L^2}J_{33}, \label{eq: J_11_ode}\\
   & \ddot{J}_{13} +\kappa^2 J_{13}=2C_x-\frac{\hat{T}_0}{J^\gamma L^2}J_{31}, \label{eq: J_13_ode}\\
   & \ddot{J}_{31} +\nu^2 J_{31}=-\frac{\hat{T}_0}{J^\gamma L^2}J_{13}, \label{eq: J_31_ode}\\
   & \ddot{J}_{33} +\nu^2 J_{33}=\frac{\hat{T}_0}{J^\gamma L^2}J_{11}. \label{eq: J_33_ode}
\end{align}
Here, the shear rate has been eliminated in favour of the radial epicyclic frequency $\kappa = \sqrt{2\Omega(2\Omega-S)}$ and we have introduced the constants $C_x$ and $C_z$ which encapsulate the conservation of circulation. These equations have the form of simple harmonic oscillators which are coupled via the pressure terms on the right hand side, facilitating the interesting dynamics present in warped systems. Each component can be endowed with some physical intuition to aid our interpretation. $J_{11}$ and $J_{33}$ represent radial and vertical stretching and compression of the ring respectively. Oscillations in these components preserve the symmetry about the midplane and correspond to breathing modes. Meanwhile $J_{31}$ represents the radial tilting and is proportional to the warp amplitude $|\psi|$. Indeed, within our local model, a stationary, global warp will pass by at the orbital rate such that the distorted geometry manifests itself as an oscillating midplane. In fact, local oscillations which are slightly detuned from the orbital frequency can be identified as slowly precessing global modes of the ring and matched onto the theory of propagating bending waves (see FOA and \cite{LubowOgilvie2000}). At orbital phases where the ring is maximally inclined, the gradient of the tilted midplane gives the warp amplitude $|\psi| = J_{31}/J_{11}$. A tilted ring results in radial pressure gradients which force sloshing motions along $x$. It is these internal oscillatory flows, driven by the warp, which are the key ingredient governing the distorted dynamics. This motion is encapsulated by $J_{13}$ which describes the vertical shearing of radial flows, such that $\mathrm{d} u_x /\mathrm{d}z \approx \dot{J}_{13}/J_{33}$. For further exposition on the interpretation of these variables the reader is referred to FOA.
%===========================================%
\section{Numerical setup}
\label{section:numerical}
%===========================================%

In FOA we made use of the grid based code \texttt{PLUTO} to validate our ring model framework in the small amplitude linear regime. However, for larger oscillations the dynamic elliptical geometry isn't compatible with fixed Cartesian boundary conditions. Indeed, for a polytropic ring, the disc matches onto a vacuum which is not well handled by grid based codes. Even in the special isothermal case, where the density smoothly tapers off, the warping motion will generate flows which pass through the domain edges, where no obvious symmetries are available to prescribe some simple or stable boundary condition. 
 
Instead we will appeal to particle based codes which naturally lend themselves to the Lagrangian framework introduced in Section \ref{section:ring_model} and avoid the worries associated with troublesome boundaries. One obvious candidate is to employ a smoothed particle hydrodynamic (SPH) code. These have been used to great effect in previous studies of warped dynamics, but might present an anomalously high diffusion due to the artificial shock capturing viscosity. Recent years have seen the development of complementary `moving-mesh' codes which aim to combine the advantages of both grid and particle schemes. Indeed, the particles' positions define an adaptive grid at each time-step. Here we employ one such code called \texttt{GIZMO}, developed by \cite{Hopkins2015}. This code descends from the \texttt{GADGET} SPH code, but crucially implements a range of distinct solution schemes. In this work we will make use of the \textit{Meshless Finite-Mass} (MFM) method which shows better conservation properties \citep{Hopkins2015,DengEtAl2017} owing to its Godunov-type numerical construction.

In this section we will describe how our numerical setup is realised in \texttt{GIZMO}. We will test this by first performing a code validation run, initialised with a small warp such that no instability is excited. Finally we will outline the main tilting mode setup, which is the focus of this paper.

%-------------------------------------------%
\subsection{Initialising the reference ring}
\label{subsection: reference_ring}
%-------------------------------------------%

\begin{figure*}
    \centering
    \includegraphics[width=\textwidth]{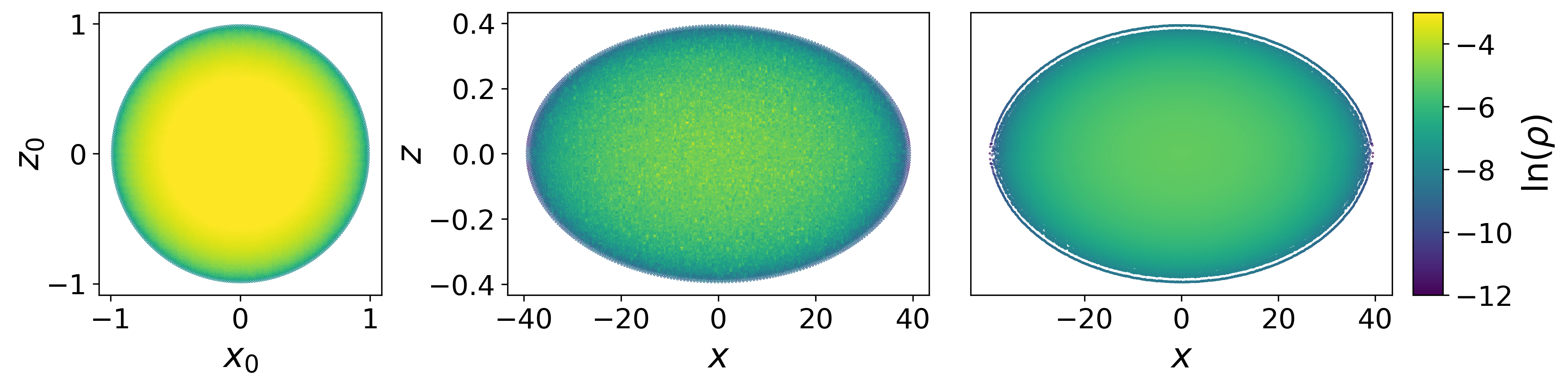}
    \caption{The particle placements in setting up the \texttt{GIZMO} equilibrium configuration. The colour bar denotes the logarithmic measure of the density which decreases from the centre of the ring outwards. \textit{Left panel}: The relaxed, stationary reference ring is mapped into the \textit{Middle panel}: noisy thin ring with $\epsilon = 0.01$. This is then frictionally relaxed for a further 60 time units to produce the smooth equilibrium state seen in the \textit{Right panel}. Note that the stretched aspect ratio exaggerates an apparent fringing of particles at the outer edge in this `glass-like' state.}
    \label{fig:equilibrium_ring_structure}
\end{figure*}

The first step is to establish a thin, steady-state ring upon which we can introduce a warp. We begin by setting up a steady circular reference state as per the Lagrangian framework described above. In order for such a state to be stationary in the code, we must artificially set a potential with circular symmetry in the $(x_0, z_0)$ poloidal plane. Taking 
\begin{equation}
\label{eq:reference_potential}
    \Phi_0 = \frac{\nu^2}{2}(x_0^2+z_0^2) = \frac{\nu^2 L^2 R^2}{2}
\end{equation}
yields a radially directed gravitational acceleration such that the hydrostatic balance is described by
\begin{equation}
\label{eq:reference_hydrostatic}
    \frac{1}{L}\frac{\partial p_0}{\partial R} = - \nu^2 L R \rho_0.
\end{equation}
The ring model framework requires that the pressure gradient term $\nabla p/\rho$ be linear in the coordinates, which imposes the condition
\begin{equation}
    \label{eq: p_rho_R_relation}
    \frac{d\Tilde{p}}{dR}=-R\Tilde{\rho},
\end{equation}
linking the pressure and density structure in the reference state. Inserting this condition into equation \eqref{eq:reference_hydrostatic} shows that the equilibrium is satisfied provided $\hat{p}_0 = \hat{\rho}_0 \nu^2 L^2$ which sets the characteristic temperature $\hat{T}_0 = \hat{p}_0/\hat{\rho}_0 = \nu^2 L^2$. Indeed, a wide range of pressure-density structures satisfy equation \eqref{eq: p_rho_R_relation}. In this work we will adopt a polytropic relationship, characterised by some index $n > 0$ and dimensionless constant $\tilde{K}$, such that $\tilde{p} = \tilde{K} \tilde{\rho}^{1+1/n}$. Integrating equation \eqref{eq: p_rho_R_relation} gives
\begin{align}
    & \tilde{\rho} = \tilde{\rho}_0\left(R_0^2-R^2 \right)^n, \label{eq:polytropic_density}\\
    & \tilde{p} = \tilde{p}_0\left(R_0^2-R^2\right)^{n+1},
\end{align}
for $R < R_0$, where
\begin{equation}
    \tilde{\rho}_0 = \left[\frac{1}{2\tilde{K} (n+1)} \right]^{n}
\end{equation}
and $\tilde{p}_0 = \tilde{K} \tilde{\rho}_0^{1+1/n}$. Thus at $R=R_0$ the density drops to 0 and the ring matches onto the surrounding vacuum. With these profiles in hand we can exploit the definition of $L^2 \equiv \frac{1}{M}\int x_0^2 \, dm$ in order to find a condition on the dimensionless radius $R_0$. In polar coordinates $x_0 = L R \cos\theta$ such that
\begin{equation}
    L^2 = L^2 \frac{\int_0^{2\pi}\int_0^{R_0}\left(R_0^2-R^2\right)^n R^3\cos^2{\theta} \,dR\,d\theta}{\int_0^{2\pi}\int_0^{R_0}\left(R_0^2-R^2\right)^nR\,dR\,d\theta},
\end{equation}
so upon cancelling the $L$ factors and computing the integrals, this dictates an expression for $R_0$,
\begin{equation}
    R_0 = \sqrt{4+2n}.
\end{equation}
Choosing the polytropic index $n$ uniquely sets $R_0$, whilst the arbitrary choice of length scale determines the characteristic temperature $\hat{T}_0$. Together these two variables tune the thermodynamic structure of our ring. We will choose $L$ such that the dimensional radius of our reference ring is unity, therefore
\begin{equation}
    L = \frac{1}{\sqrt{4+2n}}.
\end{equation}
According to our ring model, the energy equation is simply encapsulated by an adiabatic evolution with the specific entropy constant on each particle. However, we find in our preliminary simulations that the reference ring tends to puff up near the outer layers. Indeed, the internal energy seems to rise in these regions corresponding to a spurious, numerical heating. This might be due to the sound speed tending towards zero near the edge of the polytropic ring, resulting in large Mach numbers associated with the particle velocity noise. Shock heating, enhanced by the increased artificial viscosity near the boundaries, causes entropy variation on these edge particles. To circumvent this problem we enact the homentropic flag in \texttt{GIZMO} which fixes a constant value of the specific entropy for all particles. This requires the ratio $p_0/\rho_0^\gamma \equiv A_0$ to be constant everywhere. Choosing our polytropic index such that $1+1/n = \gamma$ ensures that this holds provided
\begin{equation}
    A_0 = \frac{\hat{p}_0 \tilde{K} }{\hat{\rho}_0^\gamma} = \frac{\tilde{K}  \hat{T}_0}{\hat{\rho}_0^{\gamma-1}} = \frac{\tilde{K}  \nu^2 L^2}{\hat{\rho}_0^{\gamma-1}}.
\end{equation}
The scale-free nature of ideal disc dynamics allows us to take $\hat{\rho}_0 = 1$ and assume time units such that $\Omega = \nu=1$. We wish to set the entropic constant $A_0 = 1$ in code units, which then enforces the value $\tilde{K}  = 1/L^2$. Finally, we will adopt a typical value of $\gamma = 5/3$ for the adiabatic index in our simulations. In order to initialise this reference state in \texttt{GIZMO}, we need to place particles in such a way as to emulate the target density distribution. We do this through a Monte-Carlo deposition of $N$ particles each with equal mass 
\begin{equation}
    \label{eq:particle_mass}
    m_\mathrm{p} = \frac{M}{N} = \frac{1}{N} \int \rho\,dx_0\,dz_0.
\end{equation}
The particles are then placed according to the polar coordinate description $(x_0,z_0) = L R (\cos{\theta},\sin{\theta})$ where the angle $\theta$ is chosen from the uniform distribution between $[0,2\pi]$. Meanwhile the radial position is randomly allocated by the inverse transform sampling of the normalised density distribution. We find the cumulative distribution function $Y(R)$ to be
\begin{equation}
    Y = \textrm{CDF}(R) = \frac{\int_0^{R}\left(R_0^2-R^2\right)^n R\,dR}{\int_0^{R_0}\left(R_0^2-R^2\right)^n R\,dR} = \frac{R_0^{2(n+1)}-(R_0^2-R^2)^{n+1}}{R_0^{2(n+1)}}.
\end{equation}
Then sampling uniformly in $Y\in[0,1]$ and inverting for $R$ will give the appropriately randomised particle placements. Each particle is set to have zero velocity and specific internal energy 
\begin{equation}
\label{eq:reference_internal_energy}
    e_0 = \frac{p_0}{(\gamma-1)\rho_0} = \frac{\hat{T}_0 \gamma}{2} (R_0^2-R^2),
\end{equation}
according to the analytical equilibrium prescription. In this paper we take $N = 10^6$ particles which sets a resolution capable of capturing the small scale dynamics within the ring. In future work a full resolution study would be desirable to test the convergence of our results. We will discuss this resolution further in section \ref{subsection:resolution}. Even with this large number of particles there is some inherent shot noise which leads to a deviation from the desired form of the density profile. We enact a frictional relaxation over $60$ time units wherein the noisy velocities, responding to the shot noise, undergo the damping $\mathbf{u} \rightarrow 0.99 \mathbf{u}$ at each time-step. This allows the particles to jostle into a smooth arrangement which closely matches the analytical form within the bulk of the ring and is visualised in the particle plot shown in the left hand panel of Fig.~\ref{fig:equilibrium_ring_structure}. When this relaxed system is fully released, it remains stationary as desired for this reference equilibrium.

%-------------------------------------------%
\subsection{Stretching to equilibrium ring}
\label{subsection:stretching}
%-------------------------------------------%

Now that this reference state is set up, we need to perform the Jacobian stretching to the physical thin ring equilibrium. Just as in the analytical prescription for changing coordinates, the mass on each particle is a material quantity and is left unchanged in the transformation. Furthermore, the ring maintains symmetry about the midplane so $J_{13} = J_{31} = J_{23} = 0$. Turning towards equations \eqref{eq:J11_ode} and \eqref{eq:J33_ode}, and recalling our choice of time units such that $\Omega=\nu=1$, the stretched equilibrium conditions are given by
\begin{align}
    \label{eq:stretched_equilirbium}
    2 \dot{J}_{21}+2 S J_{11}+J_{33}^{1-\gamma}J_{11}^{-\gamma} = 0, \\
    \frac{J_{11}}{J^{\gamma}}-J_{33} = 0.
\end{align}
Combining these fixes the equilibrium ring width
\begin{equation}
J_{11,\textrm{e}} = \epsilon^{-(\gamma+1)/2\gamma} , 
\end{equation}
and the azimuthal shear
\begin{equation}
    \dot{J}_{21,\textrm{e}} = -\frac{J_{11,\textrm{e}}}{2}\left(2S+\epsilon^2\right) ,
\end{equation}
in terms of the free choice for the ring aspect-ratio, $\epsilon \equiv J_{11,\textrm{e}}/J_{33,\textrm{e}}$.

We will adopt a thin ring with $\epsilon = 0.01$ as per our previous studies in FOA and FOB such that $J_{11,\textrm{e}} = 39.8$ amd $J_{33,\textrm{e}} = 0.398$. Indeed, as discussed in these previous works, such a finite but thin ring should capture the main physics of an extended disc warp. The positions and velocities are mapped using the Jacobian variables such that $x = J_{11,\textrm{e}} \, x_0$, $z = J_{33,\textrm{e}} \, z_0$ and $v_y = \dot{J}_{21,e} \, x_0$. Meanwhile, the specific internal energy and enthalpy density transform as
\begin{equation}
    h = \frac{h_0}{J_e^{\gamma-1}} = \frac{\nu^2 H(x)^2}{2}(1-\eta^2),
\end{equation}
where $J_e = J_{11,\textrm{e}} \, J_{33,\textrm{e}}$ denotes the equilibrium Jacobian determinant and $\eta = z/H$ is introduced as the dimensionless vertical coordinate, re-scaled in terms of the radially dependent disc semithickness
\begin{equation}
    \label{eq:H}
    H(x) = \sqrt{J_{33}^2-\epsilon^2 x^2}.
\end{equation}
Note that the entropy is also materially conserved so as to maintain the homentropic setup. Finally, we must modify the analytical gravity in \texttt{GIZMO} to match our local shearing box model tidal potential $\Phi_{\mathrm{t}} = -S x^2 +\frac{1}{2}z^2$ and also implement the Coriolis terms as appearing in equation \eqref{eq:local_euler_equation}. Enacting this stretching procedure, produces an approximate elliptical ring equilibrium. However, the non-isotropic nature of the stretching introduces a bias to the inter-particle spacing. This manifests itself as noise which can be seen in the middle panel of Fig.~\ref{fig:equilibrium_ring_structure}. Once again we perform a relaxation procedure and damp the spurious motions to the desired azimuthal shear profile over 60 time units. In the right hand panel we see the final relaxed state which has a very smooth profile apart from some fringing effects which develop near the edges of the ring. This fringing arises due to the drop off in particle density which reduces the resolution and will be discussed further in section \ref{subsection:resolution}. Interior to this, no such artificial structures are observed and releasing the system with the relaxation routine switched off demonstrates a steady equilibrium.

%-------------------------------------------%
\subsection{Code Resolution}
\label{subsection:resolution}
%-------------------------------------------%

\begin{figure}
    \centering
    \includegraphics[width=0.8\columnwidth]{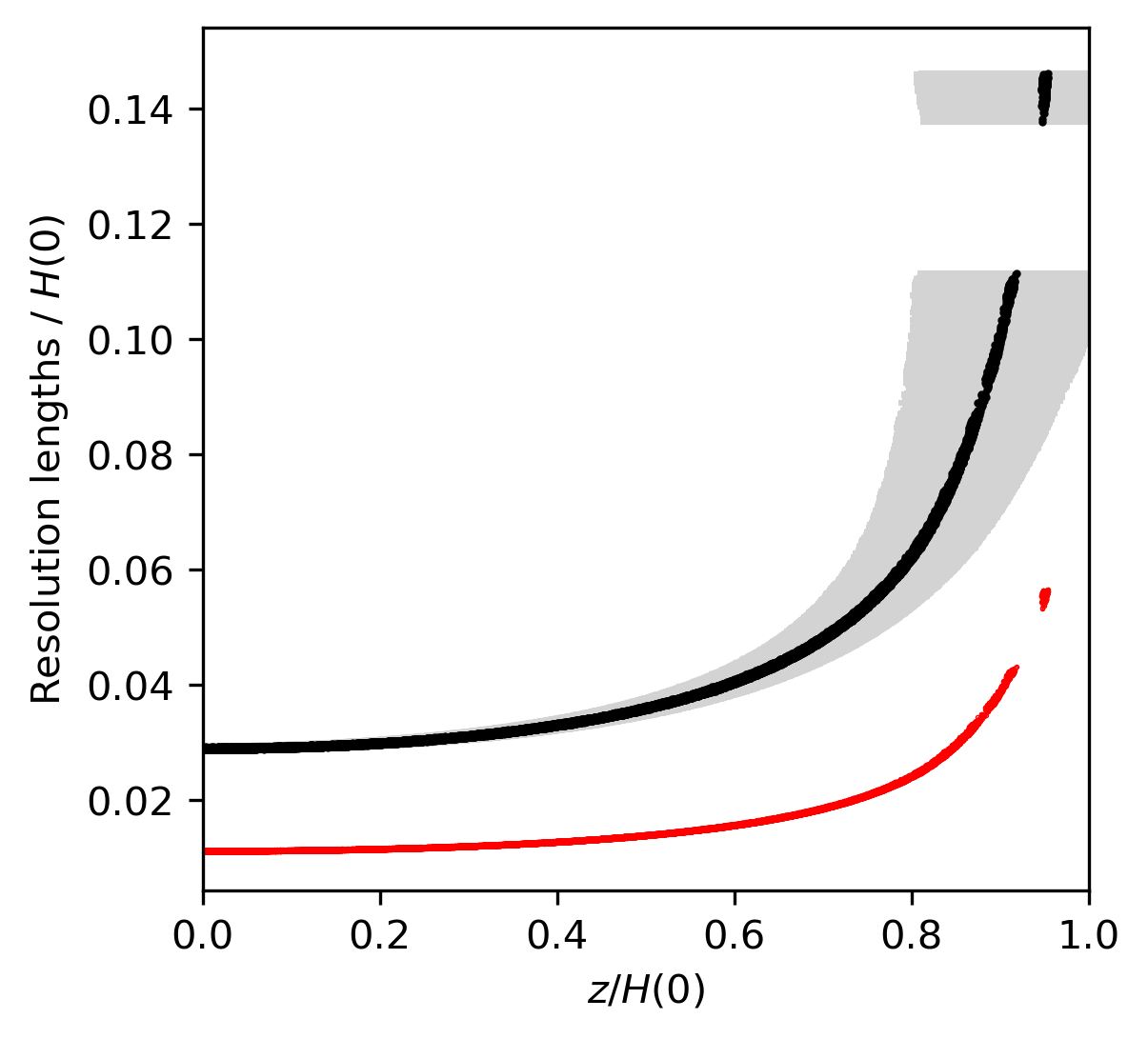}
    \caption{Vertical variation in characteristic resolution lengths for particles found within a central subdomain of the equilibrium ring $x \in [-4,4]$. All lengths are normalised with respect to the disc semithickness at the centre of the ring $H(0)$. \textit{Black points}: the kernel length $h_{\textrm{p}}$. \textit{Grey bars}: the vertical extent of each kernel's range of influence. \textit{Red points}: The inter-particle spacing.}
    \label{fig:smoothing_lengths}
\end{figure}

Lagrangian methods such as \texttt{GIZMO} rely on interpolating quantities between individual particles within a sphere of influence described by the kernel length $h_\textrm{p}$. In our simulations we employ a Wendland C4 kernel which is known to minimise numerical noise \citep{DehnenAly2012}. This adjusts itself to include 21 neighbouring particles. This number is chosen to agree with the lower end of the recommended range, as suggested by the \texttt{GIZMO} user guide\footnote{\url{http://www.tapir.caltech.edu/~phopkins/Site/GIZMO\_files/gizmo\_documentation.html}} when re-scaled to two dimensions in accordance with equation (12) in \cite{Price2012}. As the density drops near the surface, the particle resolution falls and $h_\textrm{p}$ must grow larger to encapsulate more particles. In the relaxed ring the particles repel each other to find the lowest energy state \citep{Price2011}. This results in a `glass-like' configuration as shown in the right hand panel of Fig.~\ref{fig:equilibrium_ring_structure}. Note that the apparently coherent lane of particles at the outer edge is a plotting artefact, exaggerated by the stretched aspect ratio. Upon zooming into a local patch of the ring, the particles form an approximately crystalline structure. This artefact can also be seen in Fig.~\ref{fig:smoothing_lengths} where we plot the resolution lengths in units of the disc vertical extent $H(0)$, associated with particles found within the central region of the disc between $-4<x<4$.
The kernel lengths and vertical positions of each particle are plotted as black points. The grey horizontal bars then visualise the vertical extent of each kernel. Meanwhile the inter-particle spacing is plotted as red points. The break in black and red points near the disc surface corresponds to the apparent fringing observed in the right hand panel of Fig.~\ref{fig:equilibrium_ring_structure}. Again this is an artefact which only emerges as the radial sampling width is 10 times larger than the scale height. Interior to this the resolution lengths quickly decrease towards the midplane values for the kernel length $\sim 0.03H$ and the inter-particle spacing $\sim 0.01H$. This compares favourably with previous numerical studies which have been able to capture the parametric instability in isothermal discs with the vertical scale height denoted $H_{\textrm{iso}}$. Indeed \cite{GammieEtAl2000} used a grid code with 32 cells per $H_{\textrm{iso}}$ whilst \cite{DengEtAl2020} show that a vertical resolution scale of $H_{\textrm{iso}}/8$ in the midplane is sufficient to capture the parametric instability using \texttt{GIZMO}. In accordance with \cite{DengEtAl2020}, we also find that our resolution results in an effective numerical $\alpha$ parameter of below 0.001, which is inferred from our code validation run in section \ref{subsection:code_validation} below. 

%-------------------------------------------%
\subsection{Code validation run}
\label{subsection:code_validation}
%-------------------------------------------%

We now have a thin, equilibrium ring which we wish to perturb in order to investigate the interesting warped dynamics. We do this by introducing shearing and tilting velocities which break the midplane symmetry. In practice we simply reset the poloidal velocity of each particle according to 
\begin{equation}
    \label{eq:velocity_kick}
    v_x = \frac{\dot{J}_{13}}{J_{33,\textrm{e}}} z, \quad 
    v_z = \frac{\dot{J}_{31}}{J_{11,\textrm{e}}} x.
\end{equation}
Since the initial shape of the ring is unchanged, the internal energy is the same as our equilibrium profile. 

In order to validate this setup we perform a preliminary test run which excites a low amplitude warp with small radial shear flows. In accordance with the linear theory explored in FOA, we adopt a tilting mode in equipartition with $J_{13} = J_{31} = 3.98\times10^{-3}$ such that the warp amplitude $\psi = J_{31}/J_{11,\textrm{e}} = 10^{-4}$ and the vertical shear rate $A_{13} = \mathrm{d} u_x/ \mathrm{d}z \approx \dot{J}_{13}/J_{33,\textrm{e}} = 10^{-2}$. Such a small warp amplitude will ensure that the parametric instability, which feeds off the horizontal shearing flows (see Appendix \ref{appendices:appendixA}), will have smaller growth rates and is suppressed by the numerical viscosity. Hence we expect good correspondence with the ordinary differential equations \eqref{eq:J11_ode}--\eqref{eq: J_33_ode}, which govern the dynamics within the ring model framework. Such an initial condition excites a pure bending mode with in-phase tilt and shear oscillations. The resonance between the orbital and epicyclic frequencies demands a pressure based detuning which manifests as a retrograde precession of the ring when Doppler shifted from the local orbital reference frame back into the global inertial perspective. Indeed, from linear theory the tilting mode frequency $\omega$ is predicted to be $\omega = \sqrt{1+\epsilon} \approx 1.005$. 

\begin{figure*}
    \centering
    \includegraphics[width = \textwidth]{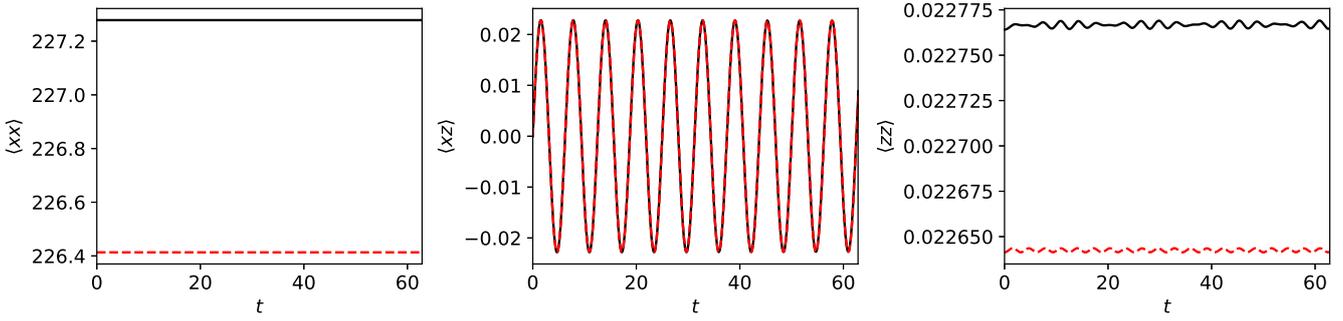}
    \caption{ The three covariance moments $\langle xx \rangle$, $\langle xz \rangle$ and $\langle ozz \rangle$ are plotted in the left, middle and right hand panels respectively over 10 orbital periods. The solid black line plots the result measured from the \texttt{GIZMO} run whilst the red dashed line is the theoretical result from our ODE framework.}
    \label{fig:code_validation}
\end{figure*}

As per the numerical study in FOA, we quantitatively test this setup by comparing the mass weighted covariance moments in our simulation with the analytical predictions computed from an implicit Runge-Kutta Radau integration of our ODE theory (see section 6.2 of FOA). In GIZMO these moments are calculated by summing over the particles according to 
\begin{equation}
    \langle x_i x_j \rangle = \frac{1}{M} \sum^{N} x_i x_j m_\textrm{p}.
\end{equation}
Meanwhile in our ODE framework these can be directly obtained as 
\begin{align}
    &\langle xx \rangle = L^2(J_{11}^2+J_{13}^2), \\
    &\langle xz \rangle = L^2(J_{11}J_{31}+J_{13}J_{33}), \\
    &\langle zz \rangle = L^2(J_{31}^2+J_{33}^2), 
\end{align}
according to equations (122)--(124) in FOA. In Fig.~\ref{fig:code_validation} the normalised covariance results for the GIZMO simulation are plotted as solid black lines whilst the ODE results provide the theoretical red dashed lines. Note that the covariance moments and other global energy statistics are output every 0.1 time units.
The middle panel is the key test since the $\langle xz \rangle$ moment measures the tilting motions which break the midplane symmetry. Over 10 orbital timescales we see very good agreement as the ring rocks back and forth with constant amplitude as expected for the initialised tilting mode. The $\langle xz \rangle$ moment exhibits only a small $\sim 0.5\%$ relative error during the early peaks. Also notice that there is a small discrepancy between the theory and simulation for the $\langle xx \rangle$ and $\langle zz \rangle$ moments. Whilst the radial and vertical moments are just constant as expected for a linear tilting mode to leading order, they have a very small fractional offset $< 1\%$ between them. This slight global modification is due to the intrinsic difficulty in defining the low density regions where there are fewer particles. Indeed, in the outer particle lane, the code density deviates by $\sim 50\%$ compared with the analytical form. Whilst the detailed numerical nature of the low resolution vacuum interface merits future attention, the dynamical consequences are clear as there is a departure from the linear flow field assumption of the ring model near the boundaries. Indeed, this numerical modification to the laminar shear flow is found in the code validation and main tilting runs (see section \ref{subsection:tilting_mode}) and should be treated with care in particle based simulations. Nonetheless, these errors can be managed in our study as we focus on the the highly resolved bulk mass content which is contained further inside the ring and dominates the dynamics. 

\begin{figure}
    \centering
    \includegraphics[width=\columnwidth]{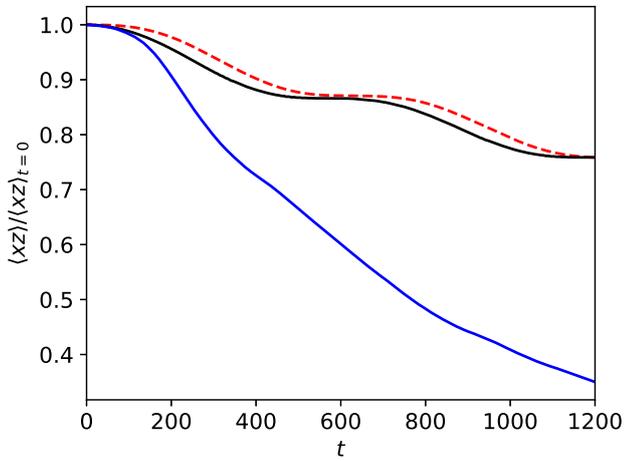}
    \caption{Fractional decay of the $\langle xz \rangle$ covariance measure versus time for different runs. \textit{Black line}: code validation simulation with $|\psi| = 10^{-4}$. \textit{Red dashed line}: the best fit to the code validation run using the viscous $\alpha$ model with $\alpha = 0.00044$. \textit{Blue line}: the main tilting run with $|\psi| = 10^{-2}$.}
    \label{fig:xz_model_comparison}
\end{figure}

Despite the decent agreement with the idealised, inviscid theory at early times, we see that the amplitude envelope of $\langle xz \rangle$ does decay over longer timescales, as indicated by the black line in Fig.~\ref{fig:xz_model_comparison}. 
Detailed examination of the small scale flow structure shows that this is not due to the emergence of any hydrodynamic instability. Attempts to extract perturbations atop the background flow (as per the methods employed later in section \ref{section:instability}) yield no growth signatures. Indeed we have specifically chosen a small warp amplitude such that there is only a small reservoir of free energy available to the parametric instability. This is in fact suppressed by a numerical viscosity which is the culprit responsible for the slow damping of tilting oscillations for this code validation run. This numerical damping can be modelled using an $\alpha$ prescription \citep{ShakuraSunyaev1973}, the details of which are contained within in Appendix \ref{appendices:appendixB}. By tuning the value of $\alpha$ we can find a good fit between the simulation and our model which is plotted as the dashed red line in Fig.~\ref{fig:xz_model_comparison}. For a value of $\alpha = 0.00044$ we get a good quantitative agreement with the fractional decay in amplitude over the extended simulation run. Furthermore we replicate the qualitative step-wise nature in which the tilt repeatedly decays and then plateaus at intervals. Since this numerical $\alpha \ll \epsilon$, the system remains comfortably within the bending wave regime (for which the system is under-damped).

%-------------------------------------------%
\subsection{Tilting mode run}
\label{subsection:tilting_mode}
%-------------------------------------------%

The main analysis of this paper will examine a fiducial tilting mode run for which we excite a larger warp amplitude which initialises stronger radial shear flows, capable of triggering the parametric instability. Motivated by the analytical tilting mode branches investigated previously in FOA and FOB, we give our equilibrium ring a velocity kick in accordance with equation \eqref{eq:velocity_kick}. Taking $\dot{J}_{13}=\dot{J}_{31} = 0.16$ corresponds to a warp amplitude $|\psi| = 4\times10^{-3}$. This is 40 times larger than our code validation run but still clearly within the linear regime. Nonetheless, the shear flow $\mathrm{d} u_x/ \mathrm{d}z = A_{13} \approx \dot{J}_{13}/J_{33,\textrm{e}} = 0.4$ is of order unity and strongly above the velocity noise level within our simulation. We expect this to be sufficient for overcoming the numerical viscosity and exciting a parametric instability. We run the simulation for 1200 time units and output full data snapshots at every unit time interval.

Before delving into the detailed analysis, it is worth commenting on the qualitative behaviour of the tilting motions, again measured by the $\langle x z \rangle$ covariance moments. Once again we find good agreement with the ODE solutions predicted by the ring model at early times. However, at later times the amplitude envelope of the $\langle xz \rangle$ moment undergoes a rapid decay, which is plotted as the blue line in Fig.~\ref{fig:xz_model_comparison}. This warp damping is faster than the code validation run and clearly differs from the numerical viscous prediction. This necessitates a detailed study of the internal flows in the disc which will be examined closely in section \ref{section:instability}. Furthermore, we find that even during early times, before any instability has appreciably grown, the flow field departs from the assumed linear form. We find that beyond about $|z|=0.2$, the $u_x$ shear flow begins to level out and turnover, owing to the reduced number of particles capable of resolving the dynamics here. Accordingly, we typically restrict our analysis domain within this region where the linear flow field is well defined.

%===========================================%
\section{Growth of parametric instability}
\label{section:instability}
%===========================================%

In this section we will closely analyse the small scale flows present in the centre of the ring. By drawing correspondence between the linear theory and simulations, we will formally identify a range of unstable wave modes which grow due to the parametric instability, before saturating and establishing a quasi-steady turbulence.

%-------------------------------------------%
\subsection{Emergence of small scale instability}
\label{subsection:instability_emergence}
%-------------------------------------------%

The emergence of small-scale structure is clearly evident in the fiducial simulation, indicating the growth of some hydrodynamic instability. In order to visualise this we zoom in on a central, localised region of the ring between $-8 < x < 8$ and $-0.2< z < 0.2$ which is chosen to avoid the artificial shear flow structures nearer the vertical boundaries where the expected linear velocity profile, an exact solution of the equations of gas dynamics, is not accurately reproduced (see previous discussion in section \ref{subsection:tilting_mode}). Note that within this region the resolution is much finer than that employed by \cite{DengEtAl2020}. Whereas they were primarily interested in capturing the emergence of the parametric instability in global simulations for the first time, our local setup allows for detailed quantitative analysis of the instability.

In accordance with the assumptions of the ring model framework, we extract the background laminar flow by finding the best fitting velocity field that is linear in the coordinates, $\bar{u}_i \equiv A_{ij}(t)x_j$. The time-dependent shear flow coefficients $A_{ij}$ are found by minimising the mass weighted sum of residuals
\begin{equation}
    \label{eq:residuals}
    \chi \equiv \frac{1}{2}\sum_{\mathcal{D}} \delta u_i \delta u_i m_\mathrm{p} = \frac{1}{2}\sum_{\mathcal{D}} (u_i-A_{ij}x_j)(u_i-A_{ik}x_k) m_\mathrm{p} , 
\end{equation}
where $\delta u_i$ denotes the residual (or perturbation) velocity components and the sum is taken over all particles in the chosen domain $\mathcal{D}$, each with mass $m_\mathrm{p}$. Similarly for the enthalpy, we find the best quadratic fit in the coordinates. We subtract this background solution to isolate the residual perturbations. For ease of analysis later, these are then interpolated from the unstructured particle mesh onto a regular grid with $n_x = 400$ and $n_z = 100$ cells in the radial and vertical directions respectively.

\begin{figure*}
    \centering
    \includegraphics[width=\textwidth]{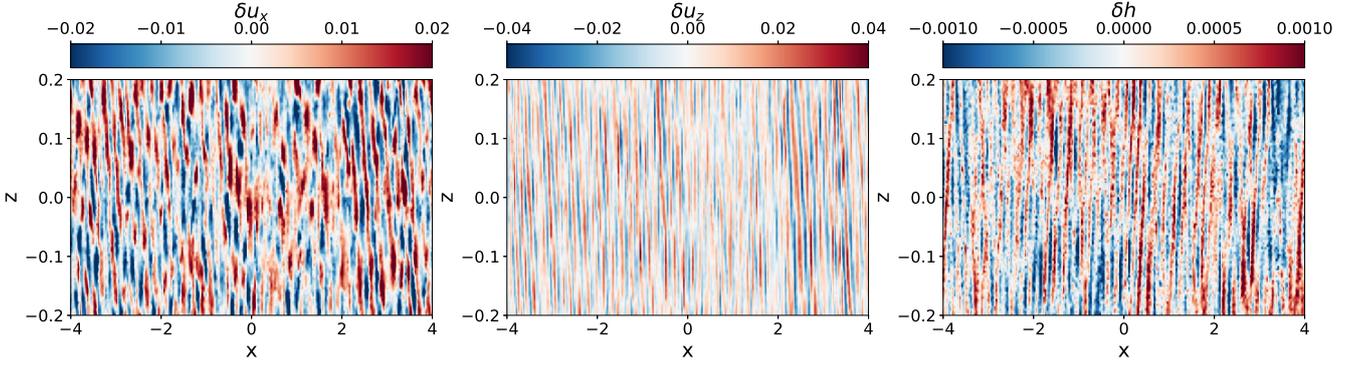}
    \caption{The perturbations are extracted from the background laminar flow according to the method described in section \ref{subsection:instability_emergence} for $t = 100$ within a central subdomain of the ring. \textit{Left panel}: The radial velocity perturbation $\delta u_x$. \textit{Middle panel}: The vertical velocity perturbation $\delta u_z$. \textit{Right panel}: The enthalpy perturbation $\delta h$.}
    \label{fig:dvx_dvz_dh_instability_100}
\end{figure*}

\begin{figure*}
    \centering
    \includegraphics[width=\textwidth]{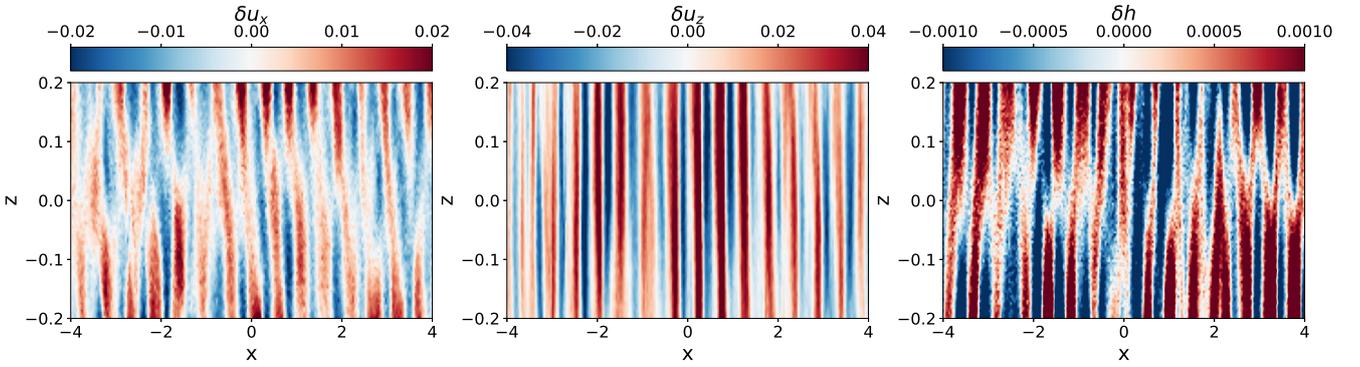}
    \caption{The same as Fig.~\ref{fig:dvx_dvz_dh_instability_100} but for $t = 400$.}
    \label{fig:dvx_dvz_dh_instability_400}
\end{figure*}

In Fig.\ref{fig:dvx_dvz_dh_instability_100} and Fig.\ref{fig:dvx_dvz_dh_instability_400} we plot the extracted radial velocity perturbations $\delta u_x$, vertical velocity perturbations $\delta u_z$, and the enthalpy perturbations $\delta h$ between $-0.4 < x < 0.4$ at two separate times, $t = 100$ and $400$. By $t = 100$ we clearly see that an instability has set in, with small scale structures in both the vertical and radial directions dominating the perturbation maps in both velocity and enthalpy. By the later time of $t=400$ the smaller scale turbulence seems to have dissipated and instead a longer radial and vertical length scale emerges as the favoured mode. This takes the form of an organised, elongated banded pattern which is reminiscent of the parametric instability predicted by linear analysis and found in previous local simulations \citep{GammieEtAl2000,OgilvieLatter2013b,PaardekooperOgilvie2019a}.
%
%-------------------------------------------%
\subsection{Parametric instability as three-mode coupling}
\label{subsection:parametric_instability_theory}
%-------------------------------------------%

Before formally identifying the parametric instability in our simulation, it is worth summarising the key theoretical features of this mechanism. As per the canonical model of parametric resonance, encapsulated by the Mathieu equation, an oscillation becomes resonantly unstable when some physical parameter of the background state varies at twice the natural frequency of the oscillator. In the case of our warping ring model, the background tilting supplies a geometrical variation at approximately the orbital frequency $\Omega = 1$. This then forces the radial oscillatory shear flows which are the fundamental source of free energy available to growing modes \citep{GammieEtAl2000}. 

\begin{figure}
    \centering
    \includegraphics[width=\columnwidth]{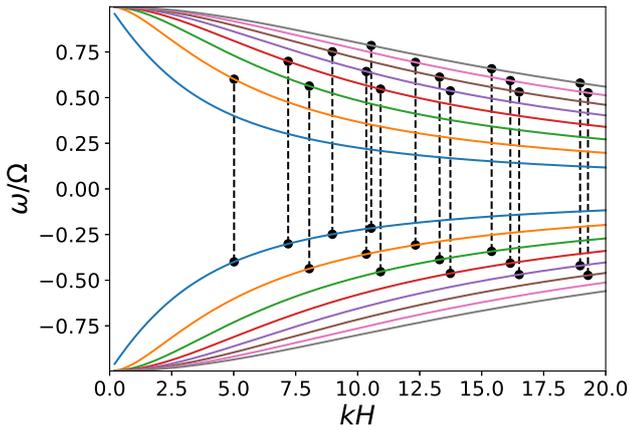}
    \caption{The dimensionless dispersion relation for a family of axisymmetric inertial modes supported by our polytropic disc. Different colours denote different vertical node numbers n (1: blue, 2: orange, 3: green, 4: red, 5: purple, 6: brown, 7: pink, 8: grey). The dashed black vertical lines of length 1 indicate the resonant couplings. These occur at wavenumbers for which two waves exist with odd and even vertical node numbers $n$ and with angular frequencies $\omega/\Omega$ and $\omega/\Omega+1$.}
    \label{fig:coupled_branches}
\end{figure}

\begin{figure}
    \centering
    \includegraphics[width=\columnwidth]{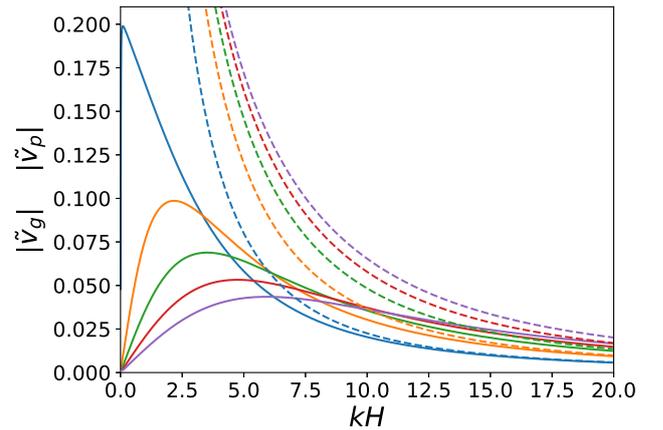}
    \caption{\textit{Solid lines}: The dimensionless group velocity of the inertial mode families. \textit{Dashed lines}: The dimensionless phase velocity of the inertial mode families. The colours label each mode according to the number of vertical nodes as per Fig.~\ref{fig:coupled_branches}.}
    \label{fig:velocity_branches}
\end{figure}

Indeed, the ring supports a wide variety of wave modes which are potential candidates for parametric instability. To find these we perturb the the local momentum and thermal equations \eqref{eq:local_euler_equation} and \eqref{eq:local_thermal_equation} about the polytropic reference disc and then linearise the system of equations. Assuming a Fourier ansatz of the form $\delta X(x,z,t) = \delta X(z) e^{i(k x-\omega t)}$ for the perturbed quantities, allows us to combine the linearised equations in favour of the enthalpy variable $\delta h$. This yields a dispersion relation given by the differential equation \eqref{eq:dispersion_relation} in Appendix \ref{appendices:appendixA}. Whilst the modes in isothermal discs (assumed by \cite{OgilvieLatter2013b}) have an analytically tractable vertical structure described by Hermite polynomials, the vertical eigenfunctions for the polytropic disc must be numerically solved as we simultaneously extract the $(\omega, k)$ eigencurves. We do this using a Chebyshev pseudo-spectral collocation method with a polynomial basis up to order 50. This gives a family of curves for different vertical eigenmodes, as shown by the solid coloured lines in Fig.~\ref{fig:coupled_branches}. Each colour denotes a different number of vertical nodes $n$.
These low-frequency branches with $|\omega | < 1$ correspond to inertial modes, as previously studied by \cite{KorycanskyPringle1995}, which are approximately incompressible perturbations restored by the Coriolis force. These curves define the dimensionless group and phase velocities, $\tilde{v}_g = (\Omega H)^{-1}\mathrm{d}\omega/\mathrm{d}k$ and $\tilde{v}_p = (\Omega H)^{-1}\omega/k$. Scaling by $\Omega H$ then yields the corresponding dimensional velocities $v_g$ and $v_p$. The magnitude of the dimensionless group and phase velocity of these modes are shown by the solid and dashed lines respectively in Fig.~\ref{fig:velocity_branches}.
Notice that due to the sloping of the dispersion relation, the group velocity and phase velocity are oppositely directed. Furthermore, the lowest order mode, plotted as the blue line, seems to exhibit a distinct limiting behaviour as $k \rightarrow 0$. In the long wavelength limit, the mode structure becomes essentially linear in the vertical coordinate and can be identified with a propagating bending mode. Indeed, it is well known that linear warping waves propagate radially at half the appropriately vertically averaged sound speed $\bar{c}$ \citep{PapaloizouLin1995}. In accordance with \cite{LubowOgilvie2000} this is given by $\bar{c} = \sqrt{P/\Sigma}$, where $P$ and $\Sigma$ are the vertically integrated pressure and density respectively. Computing this for our ring and scaling by $\Omega H$ yields $\bar{c}/(2 \Omega H) = 0.204$. This compares favourably with the limiting value of $\Tilde{v}_g \rightarrow 0.2$ for long wavelengths.

In order to destabilise these inertial waves, they must be able to communicate with each other through some mode coupling process. Following the analysis of \cite{GammieEtAl2000}, we consider the interplay between a triad of waves via weakly nonlinear interaction -- namely the background warp and two oppositely directed inertial modes with frequencies $\omega_{1}$ and $\omega_{2}$. Energy can only flow efficiently between the modes, facilitating sustained secular growth, when a resonance condition is met. Since the ring is thin, the effective radial wavenumber of the warp is 0 such that the two inertial waves should have the same value of $k$. Furthermore we demand that $\omega_2 - \omega_1 = 1$, so that the product of warped background terms with one inertial mode can resonantly force the other (and vice versa). The vertical mode structures must also satisfy a spatial coherency condition. Since the warped disturbance is odd about the midplane, the product of this with an even inertial mode demands that the other be odd. Thus only resonances between inertial branches separated by an odd number of steps can produce unstable growth. In Fig.~\ref{fig:coupled_branches} we have identified these resonant couplings as black dots connected by dashed lines. Note, that in comparison to the previous theory of three-mode resonance for an isothermal disc explored by by \cite{GammieEtAl2000} wherein only neighbouring branches can couple (see \cite{OgilvieLatter2013b} Fig.1), here we note that each branch can resonate with many others in principle and give rise to growth. This is because the product of the warped vertical structure, proportional to $z$, with the polytropic eigenfunctions, can be projected onto many other vertical modes. However, we do find that the couplings are in fact strongest for neighbouring branches and thus are expected to dominate the growth phase. Indeed, for more separated branches, there is a greater difference in the number of vertical nodes between the two inertial wave modes. Thus the overlap integral exhibits an approximate cancellation in accordance with the stationary phase approximation. 

The full three-mode coupling analysis is described in detail in Appendix \ref{appendices:appendixA} where we exploit the warped shearing box framework of \cite{OgilvieLatter2013a}. However, here we will simply summarise the main results. Consider two inertial wave packets which are supported by the background disc and have the form
\begin{equation}
    \label{eq:mode_sum_ansatz_text}
    \textbf{u}_0 = A_1(X,T)\mathbf{u}_{01}(z)e^{-i\omega_1 t}+A_2(X,T)\mathbf{u}_{02}(z)e^{-i\omega_2 t}.
\end{equation}
These two eigenmodes $\mathbf{u}_{01}$ and $\mathbf{u}_{02}$ have resonant frequencies $\omega_1$ and $\omega_2$ as described above. Furthermore, the waves have complex amplitude $A_{1}$ and $A_{2}$, encapsulating the magnitude and phase information, which can evolve slowly in space and time as the modes grow. These are governed by the evolutionary equations 
\begin{align}
    \label{eq:A_1}
    \partial_t A_1+v_{g,1}\partial_x A_1 = |\psi|C_1 A_2, \\
    \label{eq:A_2}
    \partial_t A_2+v_{g,2}\partial_x A_2 = |\psi|C_2 A_1,
\end{align}
as derived in equations \eqref{eq:A_1_evolution} and \eqref{eq:A_2_evolution} in Appendix \eqref{appendices:appendixA}. The left-hand sides simply describe the decoupled propagation of two inertial wave-packets at their respective group velocities denoted by $v_g$. Meanwhile, the right-hand sides connect the equations through the coupling coefficients $C_j$, which are calculated in equations \eqref{eq:C_j}-\eqref{eq:C_j2}. Note that the right-hand side terms are proportional to $|\psi|$, emphasising that the mode coupling is facilitated by the warped distortion and will reduce to the decoupled case when $|\psi|\rightarrow 0$. The coupling coefficients themselves crucially depend upon the mode vertical structures as well as the internal laminar shear flow amplitudes which are proportional to $U$ and $V$, as described by equations \eqref{eq:laminar_u} and \eqref{eq:laminar_v} for the radial and horizontal shearing perturbations respectively. Analysis of the linear tilting modes in the ring model, previously investigated by FOA, yields $U = 1/\epsilon = 100$ and $V = 1/(2\epsilon) = 50$ to leading order. Alternatively, we recall that the frequency for the linear tilting modes is given by $\omega \sim 1\pm \epsilon/2$. This slight detuning from the orbital rate $\Omega = 1$ can be absorbed into an effective value for $q = (3\pm \epsilon)/2$. This sets the values for
\begin{equation}
\label{eq:q_U_V_text}
U = \frac{1}{2q-3} \quad \text{and} \quad V = \frac{2-q}{2q-3} \,,   
\end{equation}
which follow from an asymptotic analysis of the laminar flows in the warped shearing box model \citep{OgilvieLatter2013a}. 

\begin{figure}
    \centering
    \includegraphics[width=\columnwidth]{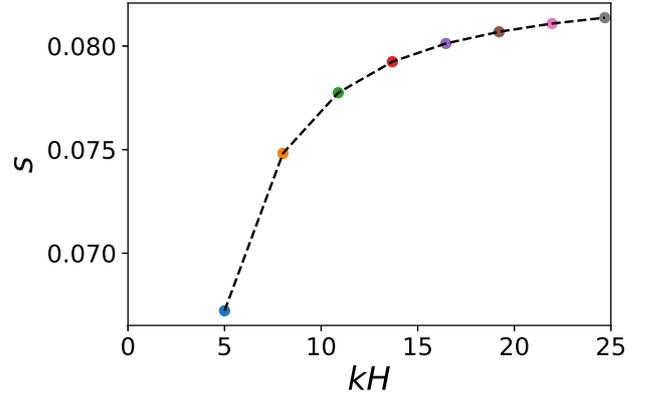}
    \caption{The theoretical growth rates predicted for the resonant couplings between neighbouring branches for $|\psi| = 10^{-2}$. The colour denotes the lowest vertical order $n$ involved in the coupling as per Fig.~\ref{fig:coupled_branches}.}
    \label{fig:growth_rates}
\end{figure}

If we insert solutions of the form $A_{j} \propto \exp{i(k_{\mathrm{w}}x - \sigma t)}$ into equations \eqref{eq:A_1} and \eqref{eq:A_2}, they can be combined into  
\begin{equation}
    \label{eq:growth_condition}
    (\sigma-k_{\mathrm{w}} v_{g,1})(\sigma-k_{\mathrm{w}} v_{g,2}) = -|\psi|^2 C_1 C_2.
\end{equation}
For an instability to occur we require that this equation has a complex conjugate pair of roots for $\sigma$ such that there is exponential growth. The solution for $\sigma$ is given by 
\begin{equation}
\label{eq:growth_rate}
\sigma = \frac{1}{2}\left( k_{\mathrm{w}}(v_{g,1}+v_{g,2}) \pm \sqrt{k_{\mathrm{w}}^2(v_{g,1}-v_{g,2})^2-4 |\psi|^2 C_1 C_2} \right).    
\end{equation}
Again we see that when $|\psi| = 0$, the resulting envelope phase velocities $\sigma/k_\mathrm{w}$ are equal to the group velocities of the two superimposed inertial mode packets. They do not interact and simply propagate away from each other. However, if we assume the inertial mode packets do not propagate and set the group velocities to be zero, then we find the maximised growth rates to be $s = i \sigma  = |\psi|\sqrt{C_1 C_2}$. The results are plotted in Fig.~\ref{fig:growth_rates} for the resonances occurring between neighbouring branches only. Indeed, we find that the couplings between neighbouring branches are strongest and hence provide the dominant growth signatures in the simulation. The colour of the points matches that of the lowest order member of the resonant pair as shown in Fig.~\ref{fig:coupled_branches}.
Here we see that the linear growth rates plateau quickly as one examines higher order couplings with shorter radial wavelengths. If we now allow for some dispersion of wave-packets, then the group velocities come back into play in equation \eqref{eq:growth_rate}. If the radial length scale associated with the disturbance is small (i.e. $k_{\mathrm{w}}$ is large) and the difference between the group velocities is large, then the first term under the square root might counteract the coupling coefficients and suppress the instability. Indeed, the parametric resonance couples branches where the group velocities are of different sign. Thus the two modes are counter-propagating which enhances this detuning effect. Physically speaking this presents a localisation of the parametric instability wherein the two interacting wave-packets pass through each other too quickly and thus don't have time to grow. For this to occur we would require that $k_{\mathrm{w}}(v_{g,1}-v_{g,2})/2> s$. For the lowest order coupling with the smallest growth rate, $s = 0.067$ and $(v_{g,1}-v_{g,2}) = 0.021$, so this requires $k_{\mathrm{w}}^{-1} < (\Delta v_g)/(2s) \sim 0.16$ which is much less than the width of the ring. Thus we anticipate that the localised nature of the growth mechanism will not be too important and expect growth rates comparable to the those plotted in Fig.~\ref{fig:growth_rates}. Equivalently we may interpret this criterion as $k_{\mathrm{w}}^{-1} < 0.4 H$, so the same result will hold for extended discs wherein the characteristic length scale of the warp is much longer than the semi-thickness.

%-------------------------------------------%
\subsection{Identifying parametric instability in the simulation}
\label{subsection:parametric_instability_simulation}
%-------------------------------------------%

\begin{figure}
    \centering
    \includegraphics[width=\columnwidth]{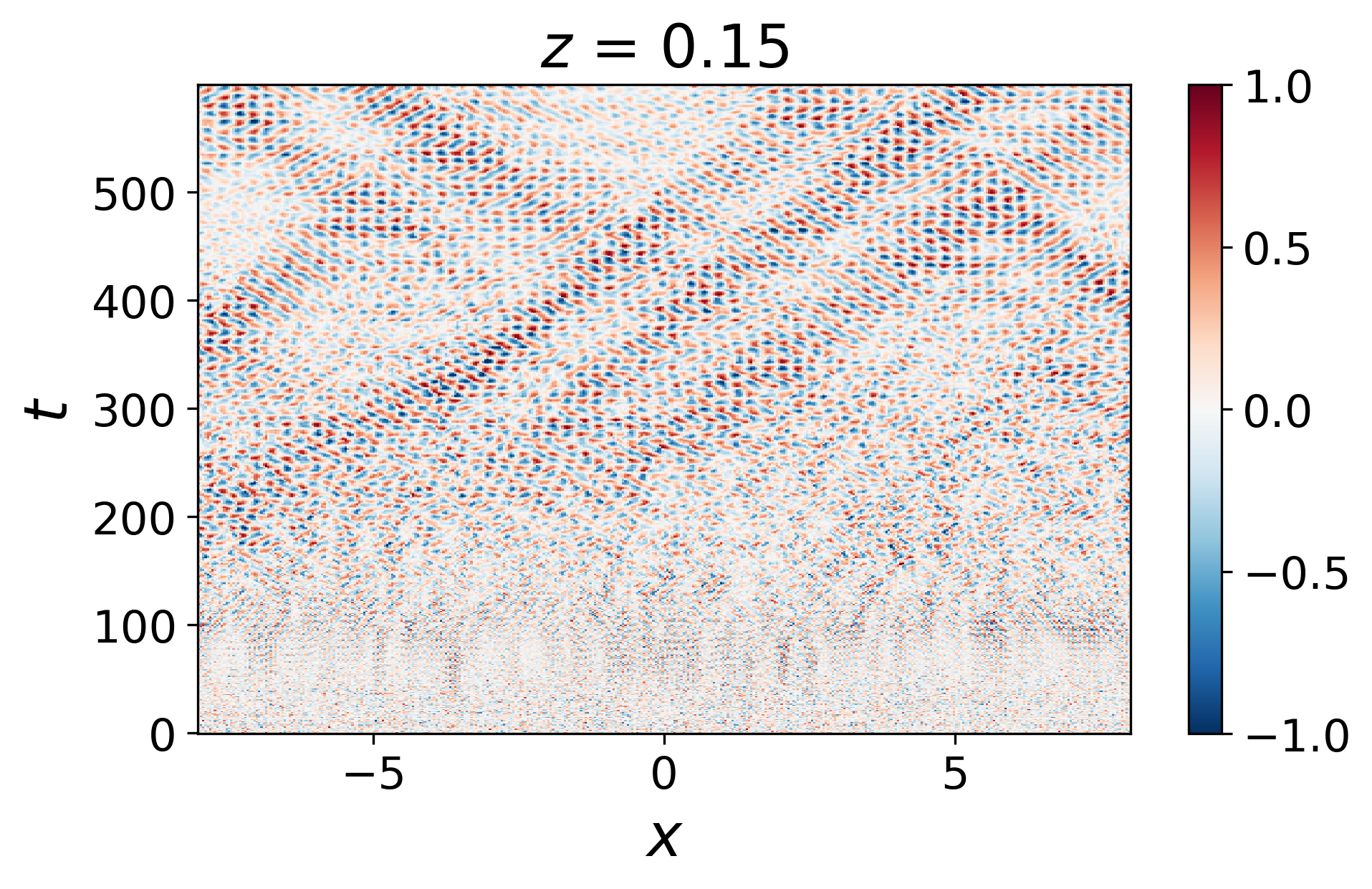}
    \caption{Horizontal slice through the data cube at $z = 0.15$ for $\delta u_z$, which has been normalised at each time-step by the maximum perturbation amplitude. Plotted for the space-time interval $-8 < x < 8$ and $0 < t < 600$.}
    \label{fig:instability_slice_dvz}
\end{figure}

Quantitative identification of this growth mechanism within a particle based code requires careful analysis. Since the parametric instability is known to manifest itself as an inertial disturbance, we expect the poloidal velocity signature to more clearly trace the instability compared with the enthalpy perturbation. Indeed, previous studies have found banded patterns in the velocity field with characteristically oblique flow fields \citep{OgilvieLatter2013b}. These present strong signatures in the vertical velocity perturbation and thus we will focus our subsequent analysis on $\delta u_{z}$. The particle perturbations are interpolated onto a grid as per the method described in Section \ref{subsection:instability_emergence} for the region $-8 < x < 8$ and $-0.2<z<0.2$. Computing this at each unit snapshot between $0 < t < 600$, extends our spatial information into a three-dimensional data cube. Before we attempt to extract the frequencies and radial wavenumbers present in the simulation, we should first normalise the data at each snapshot. Thus we divide through each time slice by the maximum value of $\delta u_z$ at that time. We can visualise this normalised data in Fig.~\ref{fig:instability_slice_dvz} where we plot $\delta u_z$ in the $x$ and $t$ dimensions for the particular vertical slice $z = 0.15$. Once again we clearly see the emergence of the instability on the smaller scales before a longer radial wavelength dominates at later times. Furthermore, we can see that beyond $t = 300$, there are small localised wave-packets which propagate left and right at the group velocity. 

\begin{figure}
    \centering
    \includegraphics[width=\columnwidth]{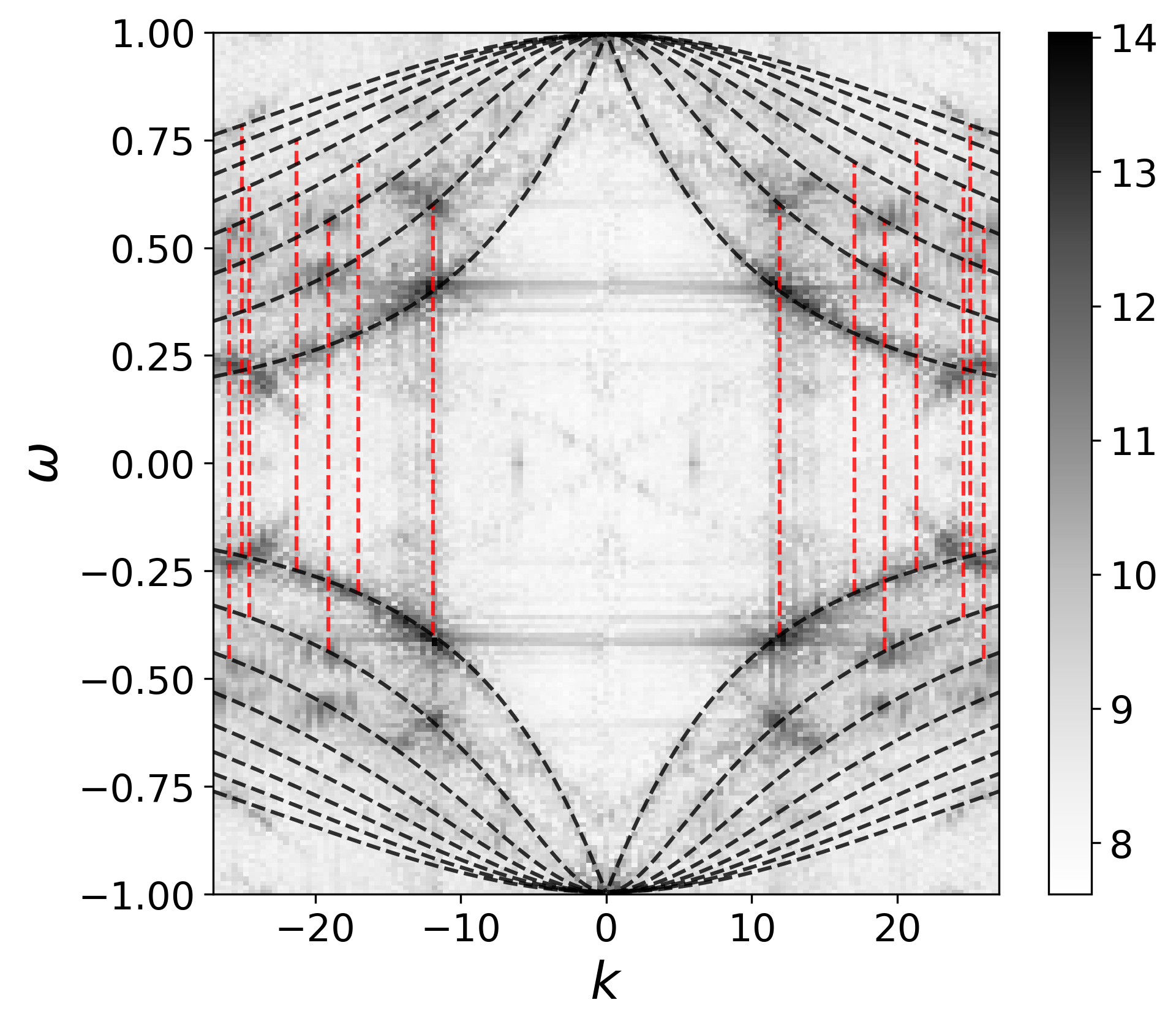}
    \caption{$\ln(1+|\widehat{\delta u_z}|)$ is plotted as the grey-scale power spectrum with dark regions indicating the presence of particular wave modes. The black dashed lines mark the theoretical dispersion relation for inertial modes, as per the dimensionless curves plotted in Fig.~\ref{fig:coupled_branches} but scaled by a factor of $H = 0.42$ for the best fit. The red dashed lines then connect the predicted resonant pair locations. Note that the figure possesses point symmetry since the signal is real. Equivalent resonances are not connected to allow for better readability of the figure.}
    \label{fig:fourier_stack_dvz}
\end{figure}

We perform a 2-dimensional Discrete Fourier Transform over the $(t,x)$ dimensions, transforming the $\delta u_z(x,z,t)$ perturbation into reciprocal $(\omega,k)$ space. The signal-to-noise is maximised by taking the absolute magnitude of this and integrating over the remaining $z$ dimension, yielding a quantity denoted $|\widehat{\delta u_z}|(k,\omega)$. The resulting 2D Fourier plane is plotted in Fig.~\ref{fig:fourier_stack_dvz}, where the colour bar denotes the value of $\ln(1+|\widehat{\delta u_z}|)$. Since the data cube is necessarily real, the complex Fourier transform exhibits $180^{\circ}$ rotational symmetry. Over-plotted are the dispersion relation branches for inertial modes and resonant couplings, as described in Section \ref{subsection:parametric_instability_theory}. Note that we have scaled the dimensionless dispersion relation derived in appendix \ref{appendices:appendixA} by $H = 0.42$ in order to achieve the best fit, which is slightly greater than the theoretical equilibrium setup with $H = 0.398$. This adjustment checks out with the slightly larger $\langle zz \rangle$ moment measured in the simulation (see Fig.~\ref{fig:code_validation}), owing to the inherent lack of particles near the vacuum boundary which slightly modifies the equilibrium, as discussed in section \ref{subsection:code_validation}. The branches are joined by red dashed lines which connect the predicted resonant locations between a pair of left and right going waves. This connectivity is reflected about the line $k=0$. Owing to the $180^{\circ}$ rotational symmetry of the Fourier map, this means they overly the physically distinct couplings for which the propagation direction of both wave modes is reversed.
We see that there is very good agreement between the predicted resonant locations and the darker patches, signifying the presence of particular modes in the simulation. The leading order coupling between the $n = (1,2)$ branches are clearly visible, as are higher order couplings at larger values of $k$. These present a degree of natural spread due to the intrinsic width of the resonance and also due to the imperfect match between the numerical model and the theoretical analysis. Such spreading faintly traces out the underlying inertial branches supported by our ring. 

\begin{figure}
    \centering
    \includegraphics[width=\columnwidth]{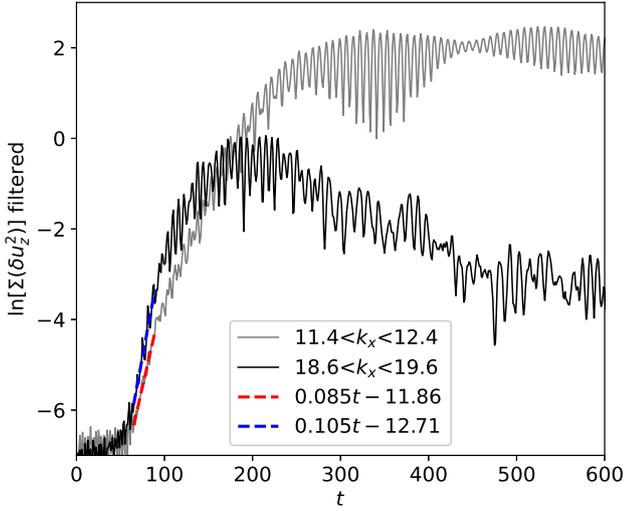}
    \caption{Growth of the vertical kinetic energy associated with resonant pairs of inertial waves. The grey and black lines show the sum of $|\delta u_z|^2$ over bands of horizontal wavenumber that include the $n=(1,2)$ and $n=(2,3)$ modes, respectively. The red and blue dashed lines denote the best exponential fit during the linear growth phase of the instability.}
    \label{fig:filtered_growth_dvz}
\end{figure}

Since there are clearly multiple different inertial waves sloshing around our simulation, it is tricky to extract a clean growth rate for a single mode. We attempt to do this by performing a 1D Fourier transform along $x$ for our unnormalized data cube. We then mask this to focus on thin windows about the resonant locations we wish to probe. Then inverse-Fourier transforming back to real space reconstructs the perturbation patterns with only the desired wavelength contributions. This `Fourier filter' allows us to track the growth of each individual component and better understand the emergence and subsequent behaviour of different length scales. In Fig.~\ref{fig:filtered_growth_dvz} we plot the natural logarithm of the summed $\delta u_z^2$ perturbations across all grid cells for two different wavenumber windows. The lighter grey line denotes the longest wavelength resonance about $11.4<k<12.4$ involving the $n = (1,2)$ branches. Meanwhile, the darker grey line denotes a shorter wavelength resonance about $18.6<k<19.6$ involving the $n = (2,3)$ branches.
Here we see that both resonant pairs undergo a clear exponential growth phase from $t \sim 65$. We have fitted straight lines to  extract the energetic growth rates during the linear onset phase of the instability between $65<t<90$, giving rates of 0.105 and 0.085 for the shorter and longer waves respectively. Halving these then gives the growth rates for the $\delta u_z$ perturbation to be $\sim 0.053$ and $0.043$. These compare reasonably well with the theoretical rates of $0.075$ and $0.067$ as calculated in Fig.~\ref{fig:growth_rates} for the two lowest order modes. Note that the slight discrepancy might originate from multiple sources. Firstly, the pipeline to extract the perturbations is rather involved and could introduce systematic errors associated with the subtraction of the background state. Furthermore, there is a damping rate $\sigma_\mathrm{d} = \alpha k^2 H^2 \Omega$, associated with the numerical viscosity $\alpha = 0.0004$ (as measured in section \ref{subsection:code_validation}). For the two lowest order modes, $k H\sim 7.5$ and $5$ which give $\sigma_\mathrm{d} \sim 0.025$ and $0.011$ respectively. These are of the order of the growth rate discrepancy, indicating that the disturbances have to battle numerical dissipation. Moreover, as the resolution drops in the lower density regions, this will also disrupt the theoretical modal structure and consequently the idealised coupling coefficients, leading to slightly suppressed growth rates. Physically, the slightly reduced growth rates could also be due to the group velocity difference as localised wave-packet regions interact for a limited time (see discussion in Section \ref{subsection:parametric_instability_theory}). However, there is no clear sign of wave-packets before $t=100$ in Fig.~\ref{fig:instability_slice_dvz} so maybe this is not a dominant effect. 

Despite these caveats, the clear exponential growth at the resonant locations strongly points towards the action of the parametric instability. After this initial rise phase, the shorter wavelength mode, tracked by the black line, appears to saturate and turnover. This can be explained by a wave breaking phenomenon wherein energy is rapidly redirected to small scales and dissipated. The kinematic wave breaking criterion requires that the crest speed of the wave exceeds the phase velocity, as has been investigated in many experimental and numerical studies of geophysical waves \citep{KhaitShemer2018}. Thus we expect saturation when $\delta u_x \sim v_{p}$. As per Fig.~\ref{fig:velocity_branches}, we see that the higher order couplings at shorter wavelengths have lower phase velocities and hence we expect these modes to break first, as confirmed by the simulation. Indeed, the resonance between the $n = (2,3)$ branches is centred around $k \sim 19.1$, for which the phase velocity magnitude of the two inertial waves is 0.023 and 0.029. Meanwhile the longest wavelength resonance between the $n = (1,2)$ branches, located around $k\sim 11.9$, involves waves with phase velocity 0.034 and 0.050.

Similar to the previous analysis, we perform a Fourier filter of the $\delta u_x$ perturbation about both of these resonances in the same localised spatial domain of the ring. First we filter about the shorter wavelength resonance taking $16.6<k<21.6$. We find that the maximum $\delta u_x$ perturbation at each time exhibits a peak of $\sim 0.014$ at around $t = 130$, coinciding nicely with the turnover of the $n = (2,3)$ resonance in Fig.~\ref{fig:filtered_growth_dvz}. Meanwhile, when filtering about $9.4<k<14.4$, the maximum $\delta u_x$ levels out about $0.025$ beyond $t = 300$. Whist these perturbation velocities are somewhat lower than the respective phase velocities, we should remember that we are only examining a localised central portion of the ring between $-0.2<z<0.2$. Indeed, the mode structure will have a larger amplitude closer to the vertical surface of the ring. Here, the density drops off and we expect higher perturbation velocities. If we repeat the above analysis, but now extend the vertical domain to $-0.3 < z < 0.3$\footnote{One should note that this extends the domain towards the region where the assumed background flow deviates from its linear form. This could introduce some systematic errors into the extraction of perturbation quantities.}, then the maximum $\delta u_x$ saturates around 0.025 for $16.6<k<21.6$ and 0.035 for $9.4<k<14.4$ which agrees very well with the predicted wave breaking amplitudes. Thus the wave breaking might be disrupting the upper regions first and then suppressing the growth in the interior domain. 

Thus, despite having the larger growth rates, the smaller scales quickly saturate and damp, whilst the longest wavelength coupled mode is left to dominate. Beyond $t = 400$ it maintains a steady amplitude as additional energy input by the parametric instability is immediately lost by continuous wave breaking. This regulates the amplitude of the wavelike turbulence and maintains it in this quasi-steady nonlinear state. The energetics of this nonlinear wave saturation process will be closely explored in the following section.
%===========================================%
\section{Modelling the turbulent feedback on the warp}
\label{section:feedback}
%===========================================%

In the previous section we identified the linear growth phase of the parametric instability and the subsequent nonlinear saturation due to wave breaking. In this section we will more carefully examine the flow of energy to the smaller scales and quantitatively investigate how the laminar warping flows evolve in response.

%-------------------------------------------%
\subsection{Energy balance}
\label{subsection:energy_balance}
%-------------------------------------------%

\begin{figure}
    \centering
    \includegraphics[width=\columnwidth]{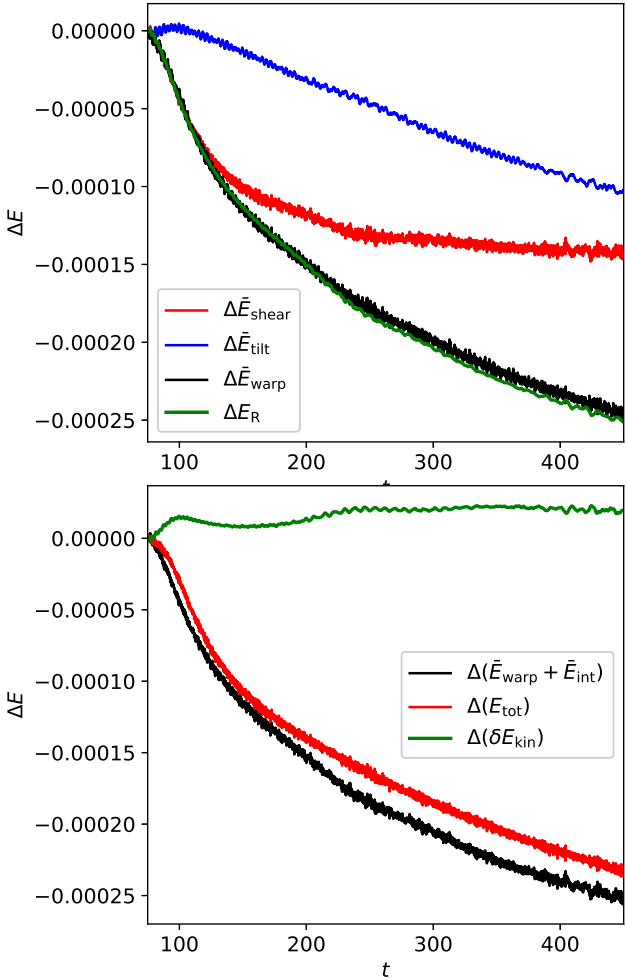}
    \caption{\textit{Upper panel}: change of energy in different components of the bulk flow from the onset of the instability saturation. Red line: horizontal shearing energy. Blue line: vertical tilting energy. Black line: total warping energy i.e. $\Delta \bar{E}_{\textrm{warp}} = \Delta \bar{E}_{\textrm{shear}}+\Delta \bar{E}_{\textrm{tilt}}$. Green line: time integrated Reynolds energy term which appears in equation \eqref{eq:energy_conservation}. \textit{Lower panel}: energy balance between the large and small scales. Black line: change in the bulk warping and internal energy. Red line: total energy in the simulation. Green line: change in perturbed kinetic energy.}
    \label{fig:energy_balance}
\end{figure}

We first need a recipe to extract the background laminar flows associated with the warp. As per our methodology in Section \ref{subsection:instability_emergence}, we find the best fitting flow field which is linear in the coordinates, $\bar{u}_i \equiv A_{ij}(t)x_j$, by minimising equation \eqref{eq:residuals}. Since we are now interested in the global energy budget of the disc, we take the summation domain $\mathcal{D}$ to include all particles. This minimisation condition imposes that $\partial \chi / \partial A_{ij} = 0$ such that
\begin{equation}
\label{eq: residual_condition}
    \sum \delta u_i x_j m_\mathrm{p} = 0.
\end{equation}
Thus the mean flow component is orthogonal to the residuals and we can show that
\begin{equation}
    \frac{1}{2}\sum u_i u_i m_\mathrm{p} = \frac{1}{2}\sum \bar{u}_i\bar{u}_i m_\mathrm{p}+\frac{1}{2}\sum \delta u_i \delta u_i m_\mathrm{p}.
\end{equation}
Here the kinetic energy is neatly partitioned between the bulk flow and perturbed motions. Furthermore, by using this specialised averaging technique we can construct an energy equation which governs the flow of kinetic energy between the large and small scales. Taking the momentum equation \eqref{eq:local_euler_equation} and decomposing the velocity field into the bulk and perturbed components, we multiply through by $\bar{u}_i$ and sum over all particles. Careful manipulation of this expression yields the global energy equation
\begin{equation}
    \label{eq:energy_conservation}
    \frac{d}{dt}\sum \left(\frac{1}{2}\bar{u}_i^2+\Phi_{\mathrm{t}}\right) m_\mathrm{p} = -\sum (\bar{u}_i \partial_i h) m_\mathrm{p} + A_{ij}R_{ij},
\end{equation}
where 
\begin{equation}
    \label{eq:reynolds_stress}
    R_{ij} = \sum \delta u_i \delta u_j m_\mathrm{p} .
\end{equation}
Note that in these equations we have employed the Einstein summation convention over the indices $i$ and $j$. The left hand side of equation \eqref{eq:energy_conservation} gives the rate of change of the mean flow kinetic energy and potential energy, whilst the first term on the right hand side contains the work done on the flow by the pressure forces. The final term involves the tensor $R_{ij}$ which can be identified as a global analogue of the Reynolds stress since it involves the product of velocity perturbations which can extract energy from the background shear flow. Furthermore, if one instead multiplies the momentum equation by $\delta u_i$ and proceeds with a similar analysis, this stress term appears again but with the opposite sign -- indicating that a Reynolds sink for the bulk flow directly feeds into the perturbation as an energy source.  
We can see this energy exchange at work in Fig.~\ref{fig:energy_balance} where the upper panel plots the difference in energy content from the onset of the instability saturation around $t = 75$. 
Here we have partitioned the energy between different bulk reservoirs. Note that, in keeping with our extracted laminar flow notation, barred variables denote the smooth background quantities whilst $\delta$ variables represent the residual perturbations. The horizontal kinetic and potential energies are grouped together as $\bar{E}_{\mathrm{shear}} =  \sum m_\mathrm{p} [\frac{1}{2}(\bar{u}_x^2+\bar{u}_y^2)-S\Omega x^2]$. Meanwhile the vertical kinetic and potential contributions are grouped as $\bar{E}_{\mathrm{tilt}} = \sum 0.5 m_\mathrm{p} (\bar{u}_z^2+\Omega^2 z^2)$. The total dynamical energy content in the warp is the sum $\bar{E}_{\mathrm{warp}} = \bar{E}_{\mathrm{tilt}}+\bar{E}_{\mathrm{shear}}$. The change in each of these quantities with time is plotted as the red, blue and black lines respectively. Here we see that the shearing motions are damped first, followed more slowly by the tilting motion. This emphasises that the parametric instability is primarily feeding off the shear, which then siphons off energy from the tilting motions as a secondary effect. The green line computes the cumulative integral of the Reynolds stresses acting on the bulk shear flows i.e. $\Delta E_R = \int_{75}^{t} A_{ij}R_{ij}dt$. Here we see that this nicely overlies the $\Delta \bar{E}_{\mathrm{warp}}$ which confirms the energy extraction from the bulk flow and its redirection towards the smaller length scales. 

In the lower panel of Fig.~\ref{fig:energy_balance}, we compare the change of energy content in both the extracted laminar flow and the residual perturbations, with the total energy dissipation in the simulation. The change in the energy associated with the laminar flow is plotted as the black line and incorporates the kinetic warping energy and the bulk internal energy, $\Delta(\bar{E}_{\mathrm{warp}}+\bar{E}_{\mathrm{int}})$. As in the upper panel, this decreases as energy is transported through the Reynolds stresses to the inertial waves. This is closely tracked by the red line which plots the change in total energy of the simulation, $E_\mathrm{tot}$, which combines both the laminar flow and small scales. This indicates that some energy is being irreversibly lost from the simulation due to numerical viscosity. Indeed, as energy is extracted from the warp and diverted to smaller scales it is more readily dissipated so we should expect approximate balance between the red and black lines. The discrepancy between the red and black lines corresponds to the energy which is stored within the kinetic perturbations, $\delta E_\mathrm{kin}$. This is plotted as the green line which exhibits a first peak around $t = 100$, associated with the saturation of the smaller wavelength resonances, and then a later plateau beyond $t = 250$, when the longer wavelengths start breaking. The energy stored in the velocity perturbations is much less than that extracted from the bulk flow. Indeed, the energy injected into these unstable modes pushes the waves beyond their breaking limit. This triggers the cascade of energy towards even smaller scales where is promptly dissipated by the code. Thus the perturbed flow is self-regulating in the sense that any added energy `overspills' and is rapidly lost. Physically speaking this wave-breaking would manifest itself as a turbulent cascade to the viscous length scale.

%-------------------------------------------%
\subsection{Viscous tilt evolution model}
\label{subsection:viscous_tilt_model}
%-------------------------------------------%

\begin{figure}
    \centering
    \includegraphics[width=\columnwidth]{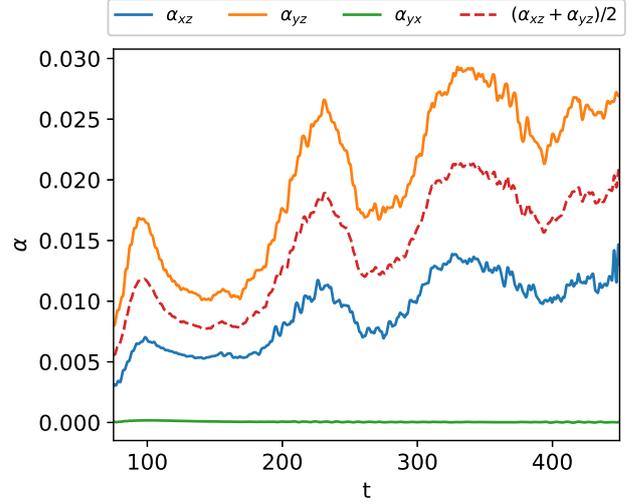}
    \caption{The time dependent value of the anisotropic $\alpha_{ij}$ coefficients which fit the proposed viscous model to the measured average Reynolds stresses. \textit{Blue line}: $\alpha_{xz}$. \textit{Orange line}: $\alpha_{yz}$. \textit{Green line}: $\alpha_{yx}$. \textit{Red dashed line}: $(\alpha_{xz}+\alpha_{yz})/2$.}
    \label{fig:alpha_time_fit}
\end{figure}

With this picture of the energy cascade and saturation process in mind, we are now well placed to develop a model for the feedback of the instability onto the warp. As is popular amongst astrophysical disc studies, such a quasi-turbulent state is often captured by treating the local Reynolds stresses as an effective viscous tensor $T_{ij}$, which is incorporated into the right hand side of the momentum equation \eqref{eq:local_euler_equation} as $\nabla\cdot \mathbf{T}/\rho$. Note that this term acts as a closure condition which encapsulates the small scales, so in accordance with our ring model, the flow fields are assumed to be laminar and linear in the coordinates. Inspired by the $\alpha$-prescription of \cite{ShakuraSunyaev1973} we will try to model each $(i,j)$ component of the local Reynolds stress tensor as 
\begin{equation}
    \label{eq:viscous_tensor}
    T_{ij} = \mu_{ij}\left(\frac{\partial \bar{u}_i}{\partial x_j}+\frac{\partial \bar{u}_j}{\partial x_i}\right) , 
\end{equation}
where this does not employ summation over the indices and  
\begin{equation}
    \label{eq:viscous_mu}
    \mu_{ij} = \frac{\alpha_{ij} \bar{p}}{\Omega} ,
\end{equation}
introduces the anisotropic $\alpha_{ij}$ parameters. Note that the symmetry condition on the viscous stress tensor, imposed by angular momentum conservation, also demands that $\alpha_{ij} = \alpha_{ji}$. Furthermore, we only consider the transport of momentum due to shearing forces such that $\alpha_{ii} = 0$. Thus we have three possible independent values for $\alpha_{xz}$, $\alpha_{yz}$ and $\alpha_{yx}$. Indeed, the strong directional dependence of the banded patterns emerging in the saturated turbulent state seen in Fig.~\ref{fig:dvx_dvz_dh_instability_400} lead us to expect different momentum transport efficiencies vertically and radially. The viscous work rate for the $i^{\textrm{th}}$ velocity contribution to the kinetic energy within our ring model is found to be
\begin{equation}
    \bar{u}_i \frac{\partial T_{ij}}{\partial x_j} = \frac{\partial}{\partial x_j}(\bar{u}_i T_{ij})-T_{ij}A_{ij},
\end{equation}
where we are now using summation notation over the $j$ index. On the right hand side, this is split up into the advective flux term minus the dissipation term. Inputting the definition of the viscous stress tensor given by equation \eqref{eq:viscous_tensor} and integrating across the ring we remove the flux term and find that the global dissipation rate for the $i^{\text{th}}$ component of the kinetic energy is given by
\begin{equation}
    \label{eq:viscous_energy_rate}
    \sum_{j} \dot{E}_{ij,\textrm{visc}} \,\, \text{with} \,\, \dot{E}_{ij,\textrm{visc}} = -(A_{ij}+A_{ji})A_{ij}\frac{\alpha_{ij}}{\Omega}(\gamma-1)\bar{E}_{\mathrm{int}}.
\end{equation}
Here we have used the result that $\int \bar{p} \, dA = \int (\gamma-1) \bar{e} \, dm = (\gamma-1) \bar{E}_{\mathrm{int}}$. This can now be compared with each $A_{ij}R_{ij}$ dissipation term derived previously in equation \eqref{eq:energy_conservation}. Convolving both of the energy dissipation rates with a unit top hat function of width 12.5 time units and then taking the ratio gives the locally smoothed fit for the time-dependent $\alpha_{ij}$ components. That is
\begin{equation}
    \alpha_{ij}(t) = - \frac{\int_{-\infty}^{\infty} R_{ij} \Pi(\tau-t)\,d\tau}{\int_{-\infty}^{\infty}(A_{ij}+A_{ji})\frac{A_{ij}}{\Omega}(\gamma-1)\bar{E}_{\mathrm{int}} \Pi(\tau-t)\,d\tau} , 
\end{equation}
where
\begin{equation}
    \Pi(\tau) = 
    \begin{cases}
        1 &  |\tau| \le \frac{\Delta\tau}{2} \\
        0 &  |\tau| > \frac{\Delta\tau}{2}
    \end{cases}
\end{equation}
with $\Delta\tau = 12.5$. The results of this are shown in Fig.~\ref{fig:alpha_time_fit}. 
Only the $\alpha_{xz}$ and $\alpha_{yz}$ coefficients are found to be significant and are plotted as the blue and orange lines respectively. Their average value is then plotted as the red dashed line. Meanwhile the $\alpha_{yx}$ coefficient remains approximately zero indicating that the parametric instability is ineffective at producing radial transport of angular momentum and the ring will undergo little spreading. A similar result was found by \cite{RyuEtAl1996} in simulations of the parametric instability in tidally distorted discs. Indeed, the turbulence arising from the parametric instability is inertial in nature which is a natural expectation since this state represents the nonlinear saturation of growing linear inertial modes (as described by the dispersion relation in Fig.~\ref{fig:coupled_branches} and the extended mode coupling analysis in appendix \ref{appendices:appendixA}). The horizontal velocity perturbations are therefore dominated by epicylic motions of fluid parcels. The radial and azimuthal velocity residuals are measured to be approximately $\pi/2$ out of phase and there is no net correlation in equation \eqref{eq:reynolds_stress}. Furthermore, the inertial behaviour suggests that the azimuthal velocity has twice the amplitude of the radial motion and hence explains why $\alpha_{yz}$ is approximately double that of $\alpha_{xz}$. The profile of the orange and blue curves exhibits a growth in the shear viscosity coefficients as the instability grows at early times. As the shortest wavelengths saturate between $t = 100-200$ we see that this overturns before the continued growth of the longest wavelength mode once again causes a rise. As the short wavelengths dissipate beyond $t = 200$ the effective $\alpha$ decreases again temporarily before undergoing a final growth phase as the longest wavelengths saturate and $\alpha$ levels out after $t = 400$. This complicated time-dependence underlines the detailed and dynamic nature of the turbulent stresses during the non-linear saturation of the instability. 
The strong anisotropy seen here demands a re-examination of the viscous theory of warped dynamics. In fact, this viscous shear can be formally incorporated in our laminar ring model as detailed in Appendix \ref{appendices:appendixB}, extending our Jacobian equation set \eqref{eq:J11_ode}-\eqref{eq:J33_ode} with the non-ideal terms  \eqref{eq:J11_visc}-\eqref{eq:J33_visc}. Assuming that the effect of this turbulent viscosity manifests itself over long timescales we are able to perform an asymptotic analysis which models the secular evolution of the shearing and tilting oscillators, $J_{13}$ and $J_{31}$ respectively. Both behave as simple harmonic oscillators at leading order with $J_{13,0} = \mathrm{Re}\left[A(T)e^{-it}\right]$ and $J_{31,0} = \mathrm{Re}\left[B(T)e^{-it}\right]$. Meanwhile $A$ and $B$ are complex shear amplitudes which are allowed to evolve over the long timescale $T = \epsilon t$. We find that these are governed by the simple coupled ODEs
\begin{align}
\label{eq:A_visc_evolution_fast}
    & d_t A = -\epsilon \frac{i}{2}B-\alpha A, \\
\label{eq:B_visc_evolution_fast}
    & d_t B = -\epsilon \frac{i}{2}A,
\end{align}
where $\alpha = (\alpha_{xz}+\alpha_{yz})/2$ is the averaged viscous coefficient plotted as the dashed red line in Fig.~\ref{fig:alpha_time_fit}. These are equivalent to equations \eqref{eq:A_visc_evolution} and \eqref{eq:B_visc_evolution} as derived in Appendix \ref{appendices:appendixB}, but simply scaled by $\epsilon$ to express in terms of the fast time coordinate. Ignoring the viscous term, these coupled equations simply correspond to a tilting mode with a precessional frequency $\omega_p = \pm \epsilon/2$. Once again, these are just the usual linear warped bending wave solutions presented in FOA. The shear viscosity then appears  as an exponential damping term on the right hand side of equation \eqref{eq:A_visc_evolution_fast} which is proportional to the vertical shear of the radial oscillation amplitude $A$. Since this damping acts on the horizontal motions, it makes sense that the effective $\alpha$ is the mean of anisotropic components $\alpha_{xz}$ and $\alpha_{yz}$. These horizontal motions are intrinsically coupled to the tilting oscillation which encapsulates the warped geometry. We will input the time dependent $\alpha$ calculated above, and numerically solve this coupled pair of ODEs to model the predicted evolution of the shear and tilt complex amplitudes. This can then be compared with the laminar values extracted from the simulation as per the method described previously in section \ref{subsection:energy_balance}. 

\begin{figure}
    \centering
    \includegraphics[width=\columnwidth]{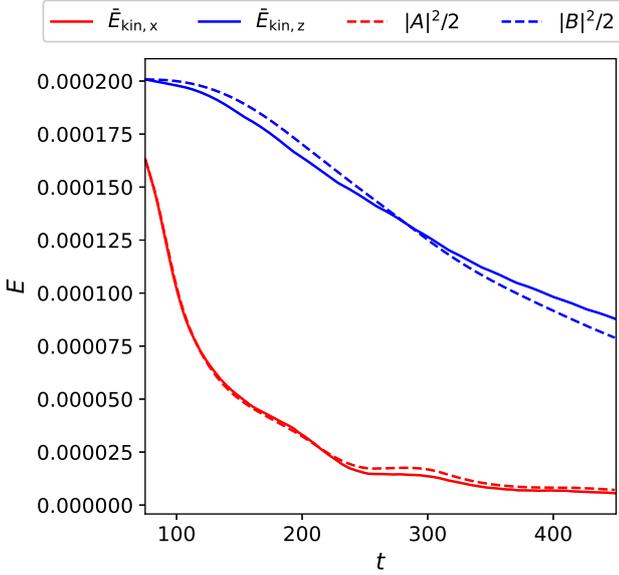}
    \caption{The comparison between the warping energy measured in the simulation (solid lines) and that predicted by the time-dependent-viscous model (dashed lines). The red lines denote the amplitude envelope for the energy contained within the radial shearing motions $\bar{E}_{\textrm{kin,x}} = 0.5\sum_{\textrm{p}} m_\textrm{p} \bar{u}_x^2$ whilst the blue lines denote the energy envelope for the vertical tilting motions $\bar{E}_{\textrm{kin,z}} = 0.5\sum_{\textrm{p}} m_\textrm{p} \bar{u}_z^2$.}
    \label{fig:viscous_tilt_model}
\end{figure}

In Fig.~\ref{fig:viscous_tilt_model} we plot the measured global radial kinetic energy envelope, which traces the secular amplitude of the shearing motions and removes the oscillations associated with the orbital timescale. This is obviously proportional to the amplitude of $J_{13}^2$ and is shown as the solid red line. Meanwhile we plot the global vertical kinetic energy envelopes, associated with the tilting motions and proportional to $J_{31}^2$, as the solid blue line. Using these to inform our initial conditions for equations \eqref{eq:A_visc_evolution_fast} and \eqref{eq:B_visc_evolution_fast}, we plot the model predictions for the energy amplitudes $|A|^2/2$ and $|B|^2/2$ as the dashed red and blue lines respectively. 
We see good quantitative and qualitative agreement between the model and simulation results. The red curves initially drop fast, indicating that the instability fundamentally sources its free energy from the shear. The blue curves then follow a slower decay as the tilt/warp responds. As a result, the shearing oscillator falls well below equipartition compared with the tilt.

\begin{figure}
    \centering
    \includegraphics[width=\columnwidth]{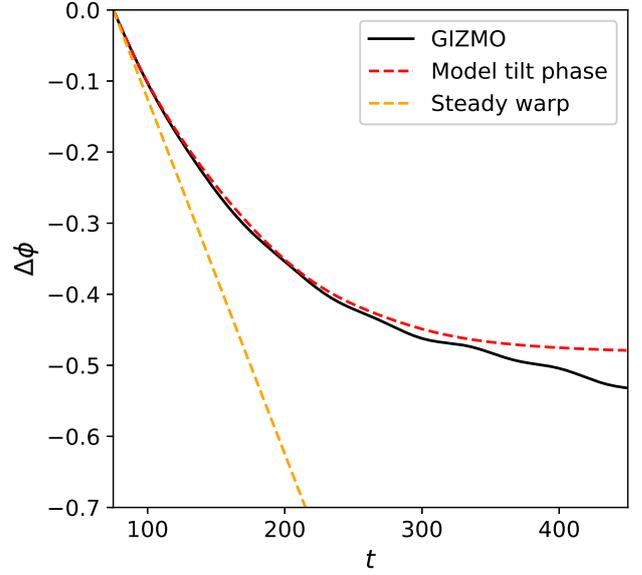}
    \caption{The difference in oscillatory phase of the $\langle x z \rangle$ covariance moment compared with the orbital phase $\Omega t$. \textit{Black line}: as measured in the simulation. \textit{Red dashed line}: as measured in the viscous warp evolution model. \textit{Orange dashed line}: the precessional phase difference predicted by the ideal ring model, which is tangential to the simulation phase evolution before the growth of the instability.}
    \label{fig:phase_evolution_time_alpha}
\end{figure}

Meanwhile in Fig.~\ref{fig:phase_evolution_time_alpha} we plot the slow phase evolution of the tilt oscillator. This is extracted from the simulation by measuring the phase difference between the covariance measure $\langle x z \rangle$ and the orbital phase $\Omega t$ through time.
This is plotted as the solid black line. Meanwhile the phase evolution predicted by the viscous ring model is shown as the red dashed line. Finally the inviscid precessional phase evolution associated with the ideal bending warping mode, is plotted as the orange dashed line. This ideal case is simply a straight line with a gradient equal to the precessional frequency which detunes the warp from the orbital frequency. We see that the effective viscosity acts to modify this behaviour. At early times, when the perturbations are still small, the phase evolves tangentially to the orange line. However, at later times this begins to flatten out as the instability grows and saturates. This indicates a decreasing precession rate as the tilting oscillations tend closer towards the orbital rate. In effect, when the local perspective of the tilting ring is Doppler shifted back into the inertial reference frame, it becomes essentially stationary whilst exhibiting only a slow damping. Now, the damping required to temper the resonance between the warp and the orbital frequency is now supplied by the turbulent viscosity. This corresponds to a transition away from the bending wave regime and into the viscous regime described by classic linear warp theory \citep{LubowOgilvie2000}. Indeed the averaged value of $\alpha \sim 0.02 \gtrsim \epsilon$ in the saturated turbulent state.
%===========================================%
\section{Discussion}
\label{section:discussion}
%===========================================%

\begin{figure}
    \centering
    \includegraphics[width=\columnwidth]{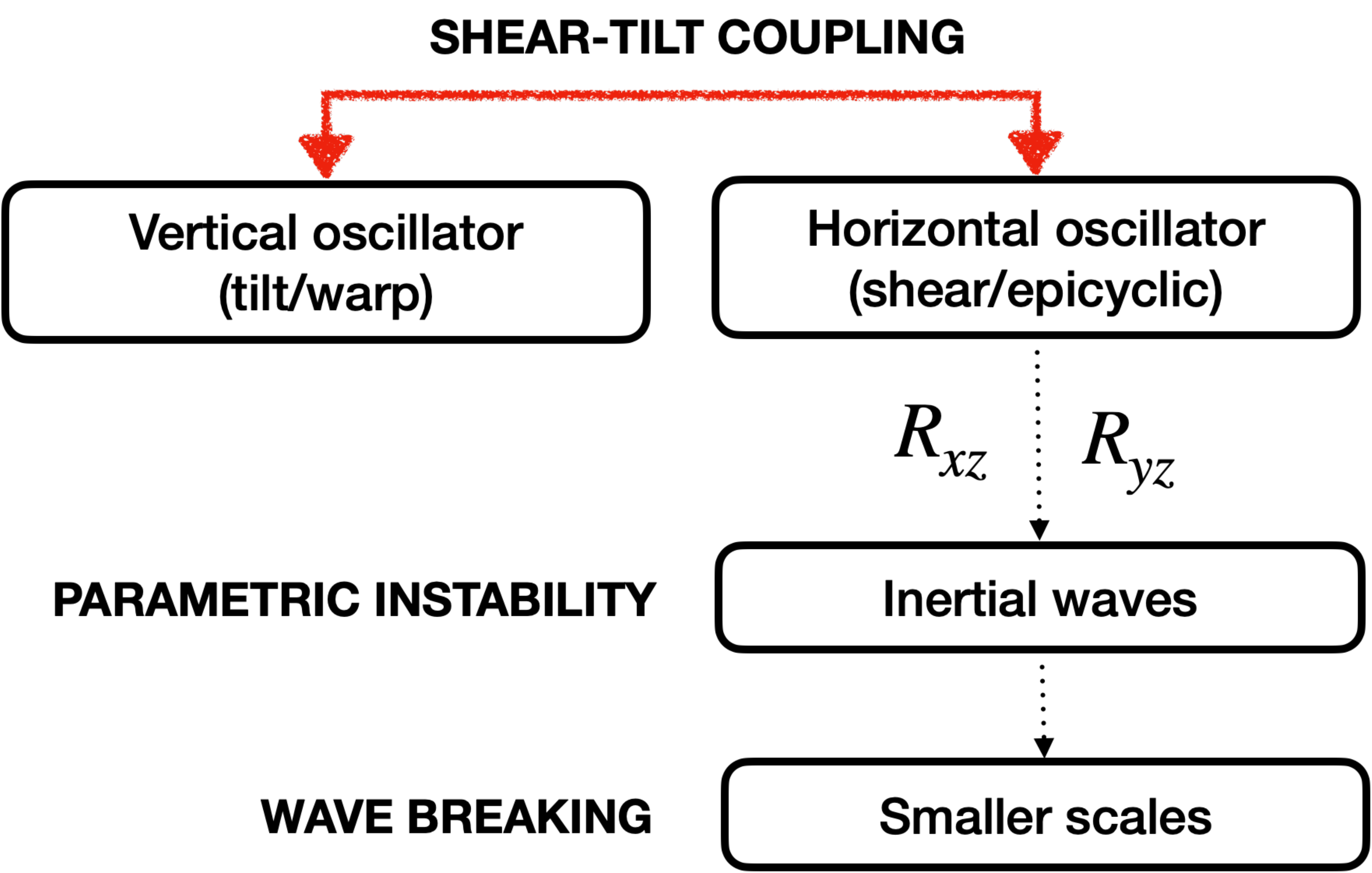}
    \caption{This schematic summarises the various couplings and energy transfer channels at play in the simulation. The two-way red arrows denote the mutual coupling between the large scale warping flows, whilst the black arrows indicate the direction of energy transfer to smaller scales. The parametric instability extracts energy from the horizontal shear oscillator which then feeds back on the warped tilting geometry through the linear shear-tilt coupling as exemplified by our viscous model. The inertial waves grow until they saturate by means of wave breaking, causing energy to cascade to the smallest scales where it is numerically dissipated. }
    \label{fig:cartoon}
\end{figure}

In this work we find that the growth and nonlinear saturation of the parametric instability provides an essential feedback onto the dynamics of a warped disc. A cartoon outlining the process is shown in Fig.~\ref{fig:cartoon} and can be summarised as follows.
The initial warp drives shearing radial motions which are obviously coupled to the azimuthal motions by the Coriolis force, making an epicyclic oscillator. This reservoir of epicyclic shearing free energy is destabilised by a three-mode coupling which excites inertial waves. These grow and saturate by means of wave breaking, at which point additional energy channelled into the inertial waves rapidly cascades to the smaller scales where it is dissipated by the numerical viscosity. This self-regulating mechanism establishes a quasi-steady turbulent state. The associated $R_{xz}$ and $R_{yz}$ Reynolds stresses dominate the transport of energy from the bulk shear flow to the inertial waves. This is effectively modelled using an anisotropic viscous $\alpha$ model\footnote{Note that our $\alpha_{ij}$ is measuring the anisotropic turbulent viscosity arising from the parametric instability and is fundamentally distinct from the $\alpha_{1}$ and $\alpha_{2}$ often appearing in other investigations of warp dynamics \citep[e.g.][]{LodatoPrice2010}.} which acts primarily on the shearing oscillator before communicating this effect to the tilting warp via a linear coupling. 

We find a range of resonant couplings growing in our simulation at radial wave numbers consistent with our polytropic coupling analysis. The linear growth rates extracted for these modes in the centre of our ring are compatible with theoretical expectations. This validates our use of the warped shearing box despite the fact it assumes an extended, horizontally homogeneous disc. This shows that the parametric instability doesn't require special periodic boundary conditions, a fixed warp or an extended structure to grow -- complementing the global findings of \cite{DengEtAl2020}. Indeed, it appears to be a robust phenomenon on local scales provided that the crossing time of inertial wave packets through the warped region (in our case the ring width $\Delta x \equiv J_{11,\textrm{e}} L$) is much longer than the growth timescale. This sets the growth criterion that $s \gg v_g/\Delta x$. This echoes the previous findings of \cite{RyuEtAl1996} who looked at the effect of boundary conditions on the growth of the parametric instability in eccentric tidally distorted discs. Here they found that the growth rate is essentially local and insensitive to the radial boundary conditions provided that the instability region is far enough away from the edges. Furthermore they find that when a  radially dependent local growth rate is introduced, the resulting global growth is reduced by the ratio of the group velocity of the inertial modes to the width of the rapid growth region. For our lowest order coupling in our main tilting run $s \Delta x/v_g \sim O(100) \gg 1$ and it is no surprise we see strong growth. However, for the code validation run the warp amplitude is a factor of $10^2$ smaller and hence so are the growth rates. This makes the instability timescale comparable to the inertial wave crossing time and hence radial propagation will contribute to the mode suppression along with the numerical viscosity. 

The saturation of the waves is found to be set by the wave breaking criterion for which their amplitude is comparable to the phase velocity. This suggests that the saturated state and associated Reynolds stresses might be independent of the initial warp amplitude, although the time taken to establish this saturated state could vary. Comparing the energy dissipation rates given by equations \eqref{eq:energy_conservation} and \eqref{eq:viscous_energy_rate}, whilst assuming $R_{ij}$ to be approximately constant, yields the scaling $\alpha_{ij} \sim 1/A_{ij} \sim 1/|\psi|$ so that the effective $\alpha$ is inversely proportional to the warp amplitude (presumably until the warp becomes so low that the underlying viscosity/noise disrupts the parametric instability mechanism and this relationship will turnover). This should be tested in future work using a range of simulations with different initial warp amplitudes. Thereafter, this prescription could be used to encapsulate the effect of the parametric instability in simplified analytical and unresolved global studies. Indeed, the importance of this mechanism is emphasised by the recent investigation of \cite{DengEtAl2020} which was the first to find the parametric instability in a global simulation. Here they find rapid damping of the warp within a few disc crossing times, reminiscent of our ring which transitions into a more diffusive regime when the instability saturates. Furthermore, our detailed local examination is the first to analyse the nonlinear feedback of this damping and its fundamentally anisotropic behaviour. This brings into question isotropic $\alpha$ prescriptions assumed in many previous theoretical and numerical studies. 

It should also be noted that the saturation process selects the longest wavelength inertial pair at later times, labelled by the vertical node count $n = (1,2)$, since the higher order resonances break and damp at lower amplitudes. These structures are highly banded and have a wavelength on the order of the disc vertical extent $H$. Whether or not this has any observational consequences demands dusty simulations of the parametric instability in warped discs and subsequent forward modelling of synthetic \textit{ALMA} images, as has been done for the vertical shear instability by \citep{BlancoEtAl2021}. This could illuminate the effect of the sub-mm dust distribution and probe kinetic information contained within molecular lines, potentially measuring the turbulence in warped regions. 

Finally we note that this simulation has been performed for a small warp amplitude, comfortably within the linear regime. However, observationally significant distortions will inevitably incur the nonlinear dynamics of warped discs. Our previous analytical efforts tried to gain a handle on such large amplitude warped dynamics and predicted the activation of strongly compressive vertical, bouncing motions twice per orbit for a tilting ring \citep{FairbairnOgilvie2021b}. It is unclear how this background flow would support the growth of the parametric instability or whether the turbulence would quickly disrupt the bouncing. This is a pertinent question for future numerical experiments. 

%===========================================%
\section{Conclusions}
\label{section:conclusion}
%===========================================%

In this paper we have performed the first detailed numerical study of the nonlinear saturation of the parametric instability in a freely evolving warped disc setup. Using a Lagrangian particle based code within a local ring model framework we observe the clear growth of different inertial modes in correspondence with the linear three-mode coupling theory for a polytropic disc. Our setup is finite in radial extent and does not enforce global coherency through periodic boundary conditions. The unstable modes grow and saturate before they have time to propagate across the ring and feel the radial boundaries. This indicates that the parametric instability is indeed a robust phenomenon capable of significant growth even for a localised warp. These modes saturate by means of wave breaking when their amplitude is similar to the phase velocity. This quenches the higher order resonances first such that the the longest wavelength inertial modes emerges as the dominant nonlinear pattern. The resultant Reynolds stresses can be modelled effectively using an anisotropic viscous $\alpha$ model which predicts a transition into a diffusive bending wave regime, highlighting that free warps in astrophysical discs are expected to damp rapidly. This suggests that some continuous misalignment must be present in observed distorted systems so as to maintain the warp.

\section*{Acknowledgements}
The authors would like to thank the anonymous reviewer for their constructive comments and suggestions. This research was supported by an STFC studentship and STFC grant ST/T00049X/1.

\section*{Data Availability}
 
Data used in this paper is available from the authors upon reasonable
request.

\typeout{}
\bibliographystyle{mnras}
\bibliography{main} 

%%%%%%%%%%%%%%%%% APPENDICES %%%%%%%%%%%%%%%%%%%%%

\appendix
%===========================================%
\section{Three-mode coupling analysis}
\label{appendices:appendixA}
%===========================================%

The linear growth phase of the parametric instability can be studied within the framework of the warped shearing box developed by \cite{OgilvieLatter2013b}. Although our experiment is performed within our freely evolving ring model to allow for a self-consistent feedback onto the warp when the instability becomes nonlinear, the warped shearing box framework should still suitably capture the linear growth phase in our simulations. The warped shearing box is radially extended and horizontally homogeneous with shearing periodic boundary conditions, so is effectively looking at a small, zoomed in portion of our thin ring. \cite{OgilvieLatter2013b} perform a three-mode coupling analysis of inertial waves in an isothermal disc and find good agreement with their numerical simulations. The isothermal disc permits neat Hermite polynomial vertical mode structures for the inertial waves, which considerably simplifies the analysis. Here we will tackle the case of a polytropic disc for which the vertical structure is more complicated. 

%-------------------------------------------%
\subsection{Warped shearing box summary}
\label{subsection:warped_shearing_box}
%-------------------------------------------%

We will begin by briefly introducing the warped shearing box which is explained in more detail in \cite{OgilvieLatter2013a}. This modifies the classical local model which is now equipped with an additional coordinate transformation
\begin{align}
    \label{eq:coordinate_transformation}
    & t^{\prime} = t, \\
    & x^{\prime} = x, \\
    & y^{\prime} = y+q\Omega t x, \\
    & z^{\prime} = z+|\psi|\cos(\Omega t)\,x,
\end{align}
where $|\psi|$ denotes the warp amplitude and $q = S/\Omega$ is the dimensionless rate of orbital shear. These primed, warped coordinates would be constant following pressureless test particles on inclined orbits. They can be substituted into the local shearing box equations \eqref{eq:local_euler_equation} and \eqref{eq:local_thermal_equation} in the case of a homentropic flow, yielding
\begin{align}
    \label{eq:warped_vx}
    & Dv_{x} - 2\Omega v_y = -\left(\partial_{x^{\prime}}+q\tau\partial_{y^{\prime}}+|\psi|\cos{\tau}\,\partial_{z^{\prime}}\right)h, \\
    \label{eq:warped_vy}
    & Dv_{y} + (2-q)\Omega v_x = -\partial_{y^{\prime}}h, \\
    \label{eq:warped_vz}
    & Dv_{z}+|\psi|\Omega \sin{\tau}\,v_x = -\Omega^2 z^\prime-\partial_{z^\prime}h, \\
    \label{eq:warped_h}
    & Dh = -(\gamma-1)h\left[\left(\partial_{x^{\prime}}+q\tau\partial_{y^{\prime}}+|\psi|\cos{\tau}\,\partial_{z^{\prime}}\right)v_x+\partial_{y^{\prime}}v_y+\partial_{z^{\prime}}v_z\right], 
\end{align}
where
\begin{equation}
    D \equiv \partial_{t^\prime}+v_x\partial_{x^\prime}+(v_y+q\tau v_x)\partial_{y^\prime}+(v_z+|\psi|\cos{\tau}v_x)\partial_{z^\prime}
\end{equation}
and $\tau=\Omega t$ is the orbital phase. The components of the relative velocity $\mathbf{v}$, after the background warp and orbital shear flow is subtracted, are given by
\begin{align}
    & v_x = u_x , \\
    & v_y = u_y+q\Omega x , \\
    & v_z = u_z-|\psi|\Omega\sin{\tau}\,x.
\end{align}

%-------------------------------------------%
\subsection{Laminar flow solutions}
\label{subsection:laminar_flow_solution}
%-------------------------------------------%
Equations \eqref{eq:warped_vx}-\eqref{eq:warped_h} admit steady state laminar solutions as the imposed warp geometry drives radial shearing flows through the oscillating pressure gradients. These follow a vertical linear shearing ansatz, 
\begin{align}
    & v_x = u(\tau) \Omega z^{\prime}, \\
    & v_y = v(\tau) \Omega z^{\prime}, \\
    & v_z = w(\tau) \Omega z^{\prime}, \\
    & h = \Omega^2[f(\tau)-\frac{1}{2}g(\tau) z^{\prime 2}]
\end{align}
echoing the linear flow field tilting modes identified within our ring model framework. This solution form will solve the warped shearing equations \eqref{eq:warped_vx}--\eqref{eq:warped_h} provided that the following ordinary differential equations are satisfied
\begin{align}
    & d_{\tau}u+(w+|\psi|\cos{\tau}\,u)u-2v = |\psi|\cos{\tau}\,g, \\
    & d_{\tau}v+(w+|\psi|\cos{\tau}\,u)v+(2-q)u = 0, \\
    & d_{\tau}w+(w+|\psi|\cos{\tau}\,u)w+|\psi|\sin{\tau}\,u = g-1, \\
    & d_{\tau}f = -(\gamma-1)(w+|\psi|\cos{\tau}\,u)f, \\
    & d_{\tau}g = -(\gamma+1)(w+|\psi|\cos{\tau}\,u)g . 
\end{align}
Following the isothermal method outlined in \cite{OgilvieLatter2013a}, one can find an asymptotic solution in powers of the warp amplitude such that the behaviour up to $\mathcal{O}(|\psi|)$ is given by
\begin{align}
   & u = |\psi|U\sin{\tau}, \label{eq:laminar_u}\\
   & v = |\psi|V\cos{\tau}, \label{eq:laminar_v}\\
   & w = 0, \label{eq:laminar_w}\\
   & g = 1 \label{eq:laminar_g}, 
\end{align}
where
\begin{equation}
\label{eq:q_U_V}
U = \frac{1}{2q-3} \quad \text{and} \quad V = \frac{2-q}{2q-3}.   
\end{equation}
Meanwhile we freely define $f\equiv H^2/2$ to be some constant at leading order, such that the equilibrium disc without any warp has the enthalpy structure
\begin{equation}
    h = \frac{\Omega^2 H^2}{2}(1-\eta^2).
\end{equation}
Here, $\eta \equiv z^{\prime}/H$ is a dimensionless vertical distance and $H$ can now be identified as the vertical extent of the disc.

%-------------------------------------------%
\subsection{Linear perturbation equations}
\label{subsection:linear_pertiurbations}
%-------------------------------------------%

We now proceed to examine disturbances atop these laminar flows by introducing the small perturbations $X\rightarrow X+ \delta X$. Using this horizontally homogeneous framework we Fourier decompose along $x$ such that perturbed quantities take the form $\delta X \propto \exp{i k x^\prime}$. Inserting this form for each quantity eventually yields the perturbation equations
\begin{align}
    \label{eq:delta_vx}
    & D \delta v_x+Bu -2 \delta v_y = -\left(ik+\partial_{x^\prime}+|\psi|\cos{\tau}\partial_{z^\prime}\right)\delta h , \\
    \label{eq:delta_vy}
    & D \delta v_y+Bv+(2-q)\delta v_x = 0 , \\
    \label{eq:delta_vz}
    & D \delta v_z+Bw+|\psi|\sin{\tau}\delta v_x = -\partial_{z^\prime}\delta h, \\
    \label{eq:delta_h}
    & D \delta h - B g z^\prime = -(\gamma-1) \left[\right.  \delta h\left(|\psi|\cos{\tau}u+w\right) \nonumber \\ 
    & \quad +h\left(ik\delta v_x+\partial_{x^\prime}\delta v_x+|\psi|\cos{\tau}\partial_{z^\prime}\delta v_x+\partial_{z^\prime}\delta v_z\right) \left.\right]
\end{align}
where
\begin{align}
    & D \equiv \partial_t+v_x\partial_{x^\prime}+i k u z^{\prime} +(v_z+|\psi|\cos{\tau}v_x)\partial_{z^\prime}, \\
    & B = \delta v_z+|\psi|\delta v_x \cos{\tau},
\end{align}
and we imposed our choice of time unit such that $\Omega =\nu = 1$ in accordance with our numerical simulations. Henceforth we will drop the primes on the transformed coordinates and derived operators for ease of notation, but one should remember that we are working within this warped reference frame. In order to analyse the evolution of these perturbations we proceed with a multiple timescale analysis. This supports the existence of waves on the fast orbital timescale, which are coupled through the warp and allowed to undergo a slow modulation over longer timescales. We will also allow for a longer length scale which can capture the envelope variation of individual wave-packets which can evolve and disperse. Thus we adopt the scaled time and distance variables $X = |\psi|x$ and $T = |\psi|t$ and introduce the expansion ansatz 
\begin{align}
    &\delta v_x = u_0(X,t,T,z)+|\psi|u_1(X,t,T,z)+\mathcal{O}{|\psi|^2} , \\
    &\delta v_y = v_0(X,t,T,z)+|\psi|v_1(X,t,T,z)+\mathcal{O}{|\psi|^2} , \\
    &\delta v_z = w_0(X,t,T,z)+|\psi|w_1(X,t,T,z)+\mathcal{O}{|\psi|^2} , \\
    &\delta h = h_0(X,t,T,z)+|\psi|h_1(X,t,T,z)+\mathcal{O}{|\psi|^2}.
\end{align}
The multiple scale variables are treated independently and so the derivative operators become
\begin{equation}
    \partial_x = |\psi| \partial_X \quad \mathrm{and} \quad
    \partial_t = \partial_0 + |\psi| \partial_1,
\end{equation}
where $\partial_0 = \partial_t$ and $\partial_1 = \partial_T$.
Inserting this form into the linearised system \eqref{eq:delta_vx}-\eqref{eq:delta_h} and expanding at each order in the warp amplitude $|\psi|$ yields a hierarchy of equations.

% ................................................%
\subsubsection{Leading order $\mathcal{O}(|\psi|^0)$:}
% ................................................%

At leading order $\mathcal{O}{|\psi|^0}$ the linearised equations give
\begin{equation}
\label{eq:zeroth_order}
\underbrace{
\begin{pmatrix}
    & \partial_0       & -2            & 0         & ik \\
    & 2-q              & \partial_0    & 0         & 0 \\
    & 0                & 0             & \partial_0 & \partial_z \\
    & i(\gamma-1)h k   & 0             & h(\gamma-1)\partial_z+\partial_z h & \partial_0 
\end{pmatrix}}_{\mathcal{L}_0}
\begin{pmatrix}
    u_0 \\
    v_0 \\
    w_0 \\
    h_0 \\
\end{pmatrix}
= \mathbf{0}
\end{equation}
where $\mathcal{L}_0$ denotes the linear matrix operator. The equations can be combined in favour of $h_0$ and, upon assuming oscillatory mode solutions $\propto \exp(-i\omega t)$, yields the leading order dispersion relation eigenvalue problem
\begin{equation}
    \label{eq:dispersion_relation}
    \mathcal{L}[\omega] h_0 = 0 , 
\end{equation}
where 
\begin{align}
\label{eq:dispersion_relation_operator}
    \mathcal{L}[\omega] \equiv & (\omega^2-\kappa^2)\left[(\gamma-1)(1-\eta^2)\frac{\partial^2}{\partial \eta^2}-2\eta\frac{\partial}{\partial \eta}\right] \nonumber \\
             & +\omega^2\left[2(\omega^2-\kappa^2)-(\gamma-1)(1-\eta^2) K^2\right],
\end{align}
is the linear differential operator which acts upon $h_0$ and we have introduced the dimensionless wavenumber $K = kH$. This has the form of a one parameter generalised eigenvalue equation for the vertical structure of $h_0$. By specifying a value for the parameter $\omega$, one can solve for a discrete set of eigenvalues $K^2$ and the corresponding eigenfunctions $h_0(\eta)$. Varying the value of the frequency $\omega$ then traces out a family of eigencurves which correspond to the dispersion branches for our linear wave modes. In order to solve this we turn towards pseudo-spectral collocation methods. Since the ring has finite boundaries at $\eta = \pm 1$ this promotes the usage of Chebyshev polynomials on a Chebyshev-Gauss root grid. The resulting dispersion relation branches are shown in Fig.~\ref{fig:coupled_branches}. Examining the region $\omega>0$, the bottom blue line represents the lowest order inertial mode with the simplest vertical structure and only $n=1$ node. The higher order modes above this exhibit an increasing number of nodes in the vertical direction.  

% ................................................%
\subsubsection{Variational Formulation}
\label{subsubsection:variational_formulation}
% ................................................%

Equation \eqref{eq:dispersion_relation} can be reinterpreted within the more general class of eigenvalue problems
\begin{equation}
    \label{eq:eigenvalue_equation}
    H \mathbf{\xi} = \lambda \mathbf{\xi},
\end{equation}
where $H$ is a Hermitian differential operator, $\lambda$ is the eigenvalue of the problem and $\xi$ is some function of the independent variable $x$. The self-adjoint nature of $H$ over some inner product space yields a range of classical techniques ripe for our disposal. In particular the problem is amenable to a variational principle. Consider the Rayleigh quotient
\begin{equation}
    \label{eq:ritz_quotient}
    \lambda = \frac{\langle \mathbf{\xi}, H \mathbf{\xi} \rangle}{\langle \mathbf{\xi}, \mathbf{\xi} \rangle},
\end{equation}
where the $\langle\cdot,\cdot\rangle$ denotes the inner product. The stationary points of this functional give the eigenvalues $\lambda$. Furthermore, we can consider what happens as the Hermitian operator itself undergoes a variation such that $H\rightarrow H+\delta H$, whilst the corresponding eigenspace is perturbed according to $\lambda\rightarrow\lambda+\delta\lambda$ and $\xi\rightarrow\xi+\delta\xi$. This will trace out a family of eigencurves which correspond to the perturbed eigenvalue equation  
\begin{align}
    & (H+\delta H)(\xi+\delta\xi) = (\lambda+\delta\lambda)(\xi+\delta\xi) \nonumber \\
    \implies & \delta H \xi+H\delta\xi=\lambda\delta\xi+\delta\lambda\xi.
\end{align}
Now taking the inner product $\langle \xi,\cdot \rangle$ of the above we find that
\begin{equation}
    \langle\xi,\delta H \xi \rangle = \delta\lambda\langle\xi,\xi\rangle,
\end{equation}
where we have used the self-adjoint property $\langle\xi,H\delta\xi\rangle=\langle H\xi,\delta\xi\rangle=\lambda\langle\xi,\delta\xi\rangle$ to cancel terms. Now assuming the variation is parameterised by $\mu$ this yields the derivative along the eigencurves to be
\begin{equation}
    \label{eq:eigencurve_derivative}
    \frac{d\lambda}{d\mu} = \frac{\langle\xi,\frac{\partial H}{\partial\mu}\xi\rangle}{\langle\xi,\xi\rangle}.
\end{equation}
Of course, in our problem the eigencurves map out the dispersion relation and the gradient is related to the group velocity of wave propagation, as we will see later. For Sturm-Liouville problems the Hermitian operator has the form
\begin{equation}
    H[\xi] = -\frac{1}{w(x)}\left[\frac{d}{dx}\left(p(x)\frac{d\xi}{dx}\right)-q(x)\xi\right]
\end{equation}
and is complemented by the inner product
\begin{equation}
    \langle f,g \rangle = \int f(x)^{*} g(x) w(x) \, dx, 
\end{equation}
over the weight function $w(x)$ and with appropriate boundary conditions, where the star denotes the complex conjugate. Inserting this into equations \eqref{eq:ritz_quotient} and \eqref{eq:eigencurve_derivative} and simplifying using integration by parts we find 
\begin{equation}
    \label{eq:sl_ritz}
    \lambda = \frac{\int p \left|\frac{d\xi}{dx}\right|^2+q\left|\xi\right|^2 dx}{\int w \left|\xi\right|^2 dx},
\end{equation}
and
\begin{equation}
\label{eq:sl_eigencurve_derivative}
\frac{d\lambda}{d\mu} = \frac{\int \frac{dp}{d\mu} \left|\frac{d\xi}{dx}\right|^2+\frac{dq}{d\mu}|\xi|^2 dx}{\int w |\xi|^2 dx}.
\end{equation}
Indeed, we can convert our second order differential equation dispersion relation into Sturm-Liouville form by multiplying equation \eqref{eq:dispersion_relation} through by the integrating factor 
\begin{equation}
    \label{eq:SL_integrating_factor}
    \frac{(1-\eta^2)^\frac{(2-\gamma)}{(\gamma-1)}}{(\omega^2-\kappa^2) (\gamma-1)}.
\end{equation}
We find that 
\begin{align}
    \label{eq:SL_form_dispersion_relation}
    -\frac{1}{(1-\eta^2)^\frac{1}{(\gamma-1)}}\left[\frac{d}{d\eta}\left(\frac{(\kappa^2-\omega^2)(1-\eta^2)^\frac{1}{(\gamma-1)}}{\omega^2}\frac{\partial h_0}{\partial\eta}\right) \right. \nonumber \\
    \left. + \frac{2(\kappa^2-\omega^2)(1-\eta^2)^\frac{(2-\gamma)}{(\gamma-1)}}{(\gamma-1)}h_0\right] = K^2 h_0 , 
\end{align}
which allows us to identify
\begin{align}
    w(\eta) & = (1-\eta^2)^\frac{1}{(\gamma-1)}, \\
    p(\eta;\omega) &= \frac{(\kappa^2-\omega^2)(1-\eta^2)^\frac{1}{(\gamma-1)}}{\omega^2}, \\
    q(\eta;\omega) &= -\frac{2(\kappa^2-\omega^2)(1-\eta^2)^\frac{(2-\gamma)}{(\gamma-1)}}{(\gamma-1)},
\end{align}
where $\omega$ is treated as the parameter which traces the dispersion branches and $K^2$ is the eigenvalue. Note that $1/(\gamma-1) = n$ is the polytropic index and so both $w$ and $p$ are proportional to the density $\rho$, as exemplified by the power law relation in equation \eqref{eq:polytropic_density}. The derivative along the eigencurves, $dK^2/d\omega = 2K dK/d\omega$,  can be computed using equation \eqref{eq:sl_eigencurve_derivative} and thus we can show
\begin{equation}
    \label{eq:analytical_dimensionless_vg}
    \tilde{v}_g = \frac{\int_{1}^{1} 2K(1-\eta^2)^\frac{1}{1-\gamma}|h_0|^2\,d\eta}{\int_{-1}^{1}\frac{4\omega}{(\gamma-1)}(1-\eta^2)^\frac{(2-\gamma)}{(\gamma-1)}|h_0|^2-\frac{2\kappa^2}{\omega^3}(1-\eta^2)^\frac{1}{(\gamma-1)}\left|\frac{\partial h_0}{\partial\eta}\right|^2 \, d\eta} ,
\end{equation}
where $\tilde{v}_g \equiv d\omega/dK$ is the dimensionless group velocity. We can perform some insightful manipulation of this result by constructing a useful integral identity from the dispersion relation. We multiply equation \eqref{eq:dispersion_relation} by $h_0$ and integrate between $-1<\eta<1$ before performing a series of integration by parts to remove the $\partial^2 h_0/\partial \eta^2$ terms. After tedious manipulations we uncover the integral identity
\begin{align}
    \label{eq:dispersion_integral_identity}
    & \int_{-1}^{1}(\omega^2-\kappa^2)(\gamma-1)(1-\eta^2)^\frac{1}{\gamma-1}\left|\frac{\partial h_0}{\partial\eta}\right|^2\,d\eta = \nonumber\\
    &\int_{-1}^{1}\omega^2\left[2(\omega^2-\kappa^2)(1-\eta^2)^\frac{2-\gamma}{\gamma-1}-(\gamma-1)(1-\eta^2)^\frac{1}{\gamma-1} K^2\right]|h_0|^2 \,d\eta.
\end{align}
Now using this identity to eliminate the term proportional to $ (1-\eta^2)^\frac{2-\gamma}{\gamma-1}$ in equation \eqref{eq:analytical_dimensionless_vg} yields 
\begin{equation}
    \label{eq:dimensionless_vg_h0}
    \tilde{v}_g = \frac{K\int_{-1}^{1}(1-\eta^2)^\frac{1}{\gamma-1}|h_0|^2\,d\eta}{\int_{-1}^{1} \frac{(\omega^2-\kappa^2)}{\omega^3}(1-\eta^2)^\frac{1}{\gamma-1}\left|\frac{\partial h_0}{\partial\eta}\right|^2 +\frac{\omega K^2}{\omega^2-\kappa^2}(1-\eta^2)^\frac{1}{\gamma-1}|h_0|^2\,d\eta}.
\end{equation}
This expression can be made more physically appealing. The individual perturbation equations in \eqref{eq:zeroth_order} yield
\begin{equation}
    h_0 = -\frac{\omega^2-\kappa^2}{\omega k} u_{0} \quad \text{and} \quad \frac{\partial h_0}{\partial \eta} = i\omega H w_{0}.
\end{equation}
Furthermore we can identify $(1-\eta^2)^\frac{1}{\gamma-1}$ as being proportional to the vertical dependence for the disc density $\rho$. Substituting these results into equation \eqref{eq:dimensionless_vg_h0} allows us to write the dimensional group velocity as
\begin{equation}
    v_g = H \tilde{v}_g = \frac{(\omega^2-\kappa^2)}{\omega k}\frac{\int \rho |u_0|^2 dz}{\int \rho\left(|u_0|^2+|w_0|^2\right) \, dz},
\end{equation}
which is in agreement with the result given by Eq. (38) of \cite{LubowOgilvie1998}.

% ................................................%
\subsubsection{First order $\mathcal{O}(|\psi|^1)$:}
% ................................................%

Now, at first order in the asymptotic equation hierarchy we find 
\begin{equation}
    \mathcal{L}_0 \mathbf{u}_1 = \mathbf{F}_1(\mathbf{u}_0)
\end{equation}
where $\mathbf{u}_i = [u_i,v_i,w_i,h_i]^T$ is the vector of perturbation quantities at the $i^{\textrm{th}}$ order and the right hand side forcing components $\mathbf{F} = [a,b,c,d]^T$, are given by

\begin{align}
\label{eq:forcing_vector}
a = & -U\sin{t} \left(i k z u_0+w_0\right)-\cos{t}\,\partial_z h_0-\partial_T u_0-\partial_X h_0 , \\
b = & -i k U z \sin{t}\,v_0 - V \cos{t}\,w_0-\partial_T v_0 ,  \\
c = & -\sin{t}\,(u_0+i k U z w_0) - \partial_T w_0 , \\
d = & -i k U z h_0\sin{t}+g z \cos{t}\,u_0-\partial_T h_0 \nonumber \\
          & -h(\gamma-1)\left(\cos{t}\,\partial_z u_0+\partial_X u_0\right).
\end{align}
Thus we see the usual inheritance of the leading operator acting on the first order perturbations on the left, which are then forced by the leading order terms on the right hand side. We can now combine the equations using the same manipulations as before and arrive at
\begin{align}
    \label{eq:1st_order_coupling}
    & h(\gamma-1)\partial_0^2\partial_z^2 h_1+\partial_z h \partial_0^2\partial_z h_1-\partial_0^4 h_1 \nonumber \\
    & \quad +\kappa^2\left[h(\gamma-1)\partial_z^2 h_1+\partial_z h \partial_z h_1-\partial_0^2 h_1\right]-(\gamma-1)h k^2 \partial _0^2 h_1 \nonumber \\
    & \quad = i(\gamma-1) h k \left(\partial_0^2 a+2 \partial_0 b\right) - \left(\kappa^2\partial_0+\partial_0^3\right)d \nonumber \\
    & \quad +\left[h(\gamma-1)\partial_0^2\partial_z+\partial_z h \partial_0^2+\kappa^2\partial_z h +\kappa^2 h (\gamma-1)\partial_z\right]c.
\end{align}
%
% ................................................%
\subsubsection{Mode coupling solvability conditions}
% ................................................%
Intuitively we understand that if the right hand side of equation \eqref{eq:1st_order_coupling} possesses any driving terms at a resonant frequency which matches the modes supported by the left hand side inertial wave operator, then there will be unbounded secular growth. Thus we must ensure that the forcing is orthogonal (over time and space) to that resonant inertial mode. Notice that the forcing components contain products of the zeroth order perturbations with unit frequency sinusoidal terms. These represent a coupling between the inertial waves with frequency $\omega$ and the background warp. Such products will generate terms with frequencies $\omega\pm 1$ which might resonantly force another inertial mode with this frequency. Therefore, without loss of generality we will consider the three-mode couplings involving the warp and inertial waves with frequency $\omega_1$ and $\omega_2 = \omega_1+1$. We construct a zeroth order solution consisting of the superposition
\begin{equation}
    \label{eq:mode_sum_ansatz}
    \textbf{u}_0 = A_1(X,T)\mathbf{u}_{01}(z)e^{-i\omega_1 t}+A_2(X,T)\mathbf{u}_{02}(z)e^{-i\omega_2 t}
\end{equation}
where $\mathbf{u}_{01}$ and $\mathbf{u}_{02}$ represent the vertical structures for the two inertial modes. These oscillate on the fast timescale according to their respective modal frequencies and are also allowed to vary on the long space and timescales according to the secular evolution of the coefficients $A_1$ and $A_2$. Other wave modes might be present in the soup of leading order noise. However these do not satisfy the resonance condition $\omega_2 = 1+\omega_1$ and hence do not lead to any secular growth. Thus they can safely be dropped from our analysis. If we insert this leading order solution into the forcing vector and extract terms corresponding to the two resonant frequencies we find that the terms proportional to $e^{-i\omega_1 t}$ are given by
\begin{align}
    \label{eq:forcing_vector_11}
    \textbf{F}_{11} \equiv 
    & \begin{pmatrix}
    a_1 \\
    b_1 \\
    c_1 \\
    d_1
    \end{pmatrix}
    = -\partial_1 
    \begin{pmatrix}
    u_{01}\\
    v_{01}\\
    w_{01}\\
    h_{01}\\
    \end{pmatrix}
    -\begin{pmatrix}
    \partial_X h_{01} \\
    0 \\
    0 \\
    h(\gamma-1)\partial_X u_{01} \\
    \end{pmatrix} \nonumber \\
    & +\begin{pmatrix}
    -\frac{1}{2}k U z u_{02}+\frac{1}{2}i U w_{02}-\frac{1}{2}\partial_z h_{02} \\
    -\frac{1}{2} k U z v_{02}-\frac{1}{2} V w_{02} \\
    \frac{1}{2} i u_{02}-\frac{1}{2}k U z w_{02} \\
    -\frac{1}{2}k U z h_{02}+\frac{1}{2} g z u_{02}+\frac{1}{2} h \partial_z u_{02}-\frac{1}{2}h\gamma\partial_z u_{02} \\ 
    \end{pmatrix}
\end{align}
whilst the terms proportional to $e^{-i\omega_2 t}$ are given by
\begin{align}
    \label{eq:forcing_vector_12}
    \textbf{F}_{12}  \equiv 
     & \begin{pmatrix}
    a_2 \\
    b_2 \\
    c_2 \\
    d_2
    \end{pmatrix}
    = -\partial_1 
    \begin{pmatrix}
    u_{02}\\
    v_{02}\\
    w_{02}\\
    h_{02}\\
    \end{pmatrix}
    -\begin{pmatrix}
    \partial_X h_{02} \\
    0 \\
    0 \\
    h(\gamma-1)\partial_X u_{02}
    \end{pmatrix} \nonumber \\
    & +\begin{pmatrix}
    \frac{1}{2}k U z u_{01}-\frac{1}{2}i U w_{01}-\frac{1}{2}\partial_z h_{01} \\
    \frac{1}{2} k U z v_{01}-\frac{1}{2} V w_{01} \\
    -\frac{1}{2} i u_{01}+\frac{1}{2}k U z w_{01} \\
    \frac{1}{2}k U z h_{01}+\frac{1}{2} g z u_{01}+\frac{1}{2} h \partial_z u_{01}-\frac{1}{2}h\gamma\partial_z u_{01}
    \end{pmatrix}.
\end{align}
 Now we assume an oscillatory ansatz for the first order response of the system forced by each frequency such that $h_1 = h_{11} e^{-i\omega_1 t}+h_{12}e^{-i\omega_2 t}$ with $j \in (1,2)$. Substituting this into equation \eqref{eq:1st_order_coupling} and rearranging gives
\begin{align}
    & \mathcal{L}[\omega_1] h_11 e^{-i\omega_1 t}+\mathcal{L}[\omega_2] h_12 e^{-i\omega_2 t}  = \nonumber \\ & F(\omega_1,\mathbf{F_{11}})e^{-i\omega_1 t}+F(\omega_2,\mathbf{F_{12}})e^{-i\omega_2 t} +\text{Non-resonant terms} ,
\end{align}
where the forcing function is given by
\begin{align}
    \label{eq:forcing_function}
    F(\omega_j,\mathbf{F}_{1j}) = 
    & 2i(\gamma-1)h k  \left(\omega_j^2 a_j+2i\omega_j b_j \right)+2i \omega_j\left(\omega_j^2-\kappa^2\right)d_j \nonumber \\
    & +2(\omega_j^2-\kappa^2)\left[h(\gamma-1)\partial_z+\partial_z h\right]c_j .
\end{align}
Now Fourier extracting the terms with equal frequencies yields two conditions for $j = (1,2)$ which must be satisfied, namely
\begin{equation}
    \label{eq:resonant_forcing_equation}
    \mathcal{L}[\omega_j] h_{1j} = F(\omega_1,\mathbf{F_{1j}}).
\end{equation}
As before, the linear operator $\mathcal{L}$ can be converted into self-adjoint Sturm-Liouville form by multiplying through by $(1-\eta^2)^\frac{2-\gamma}{\gamma-1}/w(\eta)$. The Fredholm solvability condition then imposes that eigenfunctions for the self-adjoint operator must be orthogonal with respect to the resonant forcing terms on the right hand side,
\begin{equation}
    \langle h_{0j}, \frac{(1-\eta^2)^\frac{2-\gamma}{\gamma-1}}{w(\eta)} F(\omega_j,\mathbf{F_{1j}}) \rangle = 0 ,
\end{equation}
where $h_{0j}$ is the leading order vertical eigenfunction associated with the eigenfrequency $\omega_j$. Evaluating this condition for both resonant frequencies and associated forcing functions we find the evolutionary equations
\begin{align}
    \label{eq:A_1_evolution}
    \partial_T A_1+v_{g,1}\partial_X A_1 = C_1 A_2, \\
    \label{eq:A_2_evolution}
    \partial_T A_2+v_{g,2}\partial_X A_2 = C_2 A_1,
\end{align}
where 
\begin{align}
    \label{eq:vg_C}
    v_{g,j} = \frac{C_{j2}}{C_{j1}}, \\ 
    \label{eq:C_j}
    C_j = -\frac{C_{j0}}{C_{j1}},
\end{align}
with
\begin{align}
    \label{eq:C_10}
    & C_{10} = \int_{-1}^{1} \frac{K}{2\omega_2(\omega_2^2-\kappa^2)}(1-\eta^2)^{\frac{2-\gamma}{\gamma-1}} h_{01}\times \nonumber\\
    &\quad \left\{\eta\omega_2 
        \left(-\frac{1}{2}(1-\eta^2)K^2 U(\gamma-1)\omega_1(\kappa^2+\omega_1\omega_2) \right.\right. \nonumber \\
    &\quad \left. -(\kappa^2-\omega_1^2)[(\omega_1-1)\omega_2+U\omega_1(\kappa^2-\omega_2^2)]\right) h_{02} \nonumber \\
    &\quad +\left(-U\eta^2(\kappa^2-\omega_1^2)(\kappa^2-\omega_2^2)+\frac{1}{2}(1-\eta^2)(\gamma-1)\times \right. \nonumber \\
    &\quad \left[ 2 V \omega_1(\kappa^2-\omega_2^2)+U(\kappa^2-2\omega_1^2)(\kappa^2-\omega_2^2) \right. \nonumber \\
    &\quad \left.\left. +\omega_2\left(\kappa^2(\omega_1^2+\omega_2(\omega_1-1))-\omega_1^2\omega_2(\omega_1+\omega_2-1)\right) \right] \right)\frac{\partial h_{02}}{\partial \eta} \nonumber \\
    &\quad \left. +\frac{1}{2}(1-\eta^2)U\eta(\gamma-1)(\kappa^2-\omega_1^2)(\kappa^2-\omega_2^2)\frac{\partial^2 h_{02}}{\partial\eta^2}  \right\} \, d\eta ,
\end{align}
\begin{align}
    \label{eq:C_20}
    & C_{20} = \int_{-1}^{1} \frac{K}{2\omega_1(\omega_1^2-\kappa^2)}(1-\eta^2)^{\frac{2-\gamma}{\gamma-1}} h_{02}\times \nonumber\\
    &\quad \left\{ \eta\omega_1\left(\omega_1(\omega_2+1)(\omega_2^2-\kappa^2)+U\omega_2\left[(\kappa^2-\omega_1^2)(\kappa^2-\omega_2^2) \right.\right.\right. \nonumber \\
    &\quad \left.\left. +\frac{1}{2}(1-\eta^2)K^2(\gamma-1)(\kappa^2+\omega_1\omega_2)\right]\right)h_{01} + \nonumber \\
    &\quad \left( U\eta^2(\kappa^2-\omega_1^2)(\kappa^2-\omega_1^2)-\frac{1}{2}(1-\eta^2)\left[ U(\kappa^2-2\omega_2^2)(\kappa^2-\omega_1^2) \right. \right. \nonumber \\
    &\quad +\omega_1^2\omega_2(2V+\omega_2(1+\omega_1+\omega_2)) \nonumber \\
    &\quad \left.\left. -\kappa^2(2V\omega_2+\omega_1(\omega_1+\omega_1\omega_2+\omega_2^2))\right]\right)\frac{\partial h_{01}}{\partial \eta} \nonumber \\
    &\quad \left. -\frac{1}{2}(1-\eta^2)U\eta(\gamma-1)(\kappa^2-\omega_1^2)(\kappa^2-\omega_2^2)\frac{\partial^2 h_{01}}{\partial\eta^2}\right\}\, d\eta ,
\end{align}
\begin{align}
    \label{eq:C_j1}
    & C_{j1} = -\int_{-1}^{1} \frac{(1-\eta^2)^{\frac{2-\gamma}{\gamma-1}}h_{0j}}{\omega_j(\omega_j^2-\kappa^2)}\left\{ \omega_j^2\left[ (\omega_j^2-\kappa^2)^2 \right.\right. \nonumber\\
    &\quad \left. +\frac{K^2}{2}(1-\nu^2)(\gamma-1)(\omega_j^2+\kappa^2)\right] h_{0j} \nonumber \\
    &\quad \left. +(\omega_j^2-\kappa^2)^2\left[\eta\frac{\partial h_{0j}}{\partial\eta}-\frac{(1-\nu^2)(\gamma-1)}{2}\frac{\partial^2 h_{0j}}{\partial \eta^2}\right]\right\} \,d\eta ,
\end{align}
\begin{align}
    \label{eq:C_j2}
    C_{j2} =  -H\int_{-1}^{1} K (1-\eta^2)^{\frac{1}{\gamma-1}}(\gamma-1)\omega_j^2 h_{0j}^2 \,d\eta.
\end{align}
Careful manipulation of the integral ratio $v_{g,j} \equiv C_{j2}/C_{j1}$ allows us to identify this with the group velocity of the $j^\text{th}$ resonant mode, derived previously in subsection \ref{subsubsection:variational_formulation}. 
%===========================================%
\section{Anisotropic viscous ring model}
\label{appendices:appendixB}
%===========================================%

%-------------------------------------------%
\subsection{Linear warp evolution}
\label{subsection:linear_warp_evolution}
%-------------------------------------------%

In order to identify the turbulent Reynolds stresses with an effective viscosity, it is instructive to develop evolutionary equations for the viscously damped tilt and shear within the framework of our ring model. Our simulation presents evidence for a strongly directionally dependent saturated turbulent state which suggests that we adopt the anisotropic viscous stress tensor given by equation \eqref{eq:viscous_tensor}. Appending the viscous force $\nabla\cdot\mathbf{T}/\rho$ to equation \eqref{eq:local_euler_equation} then modifies our usual ring model ODEs. The right hand side of equations \eqref{eq:J11_ode} -- \eqref{eq:J33_ode} gain the viscous terms
\begin{align}
\label{eq:J11_visc}
    \ddot{J}_{11,\mathrm{v}} = & -\frac{\hat{T}_0}{\Omega L^2 J^{\gamma+1}} \bigl( -2\alpha_{11}\dot{J}_{13}J_{31}J_{33}+2\alpha_{11}\dot{J}_{11}J_{33}^2+\alpha_{13}\dot{J}_{11}J_{13}^2-  \nonumber \\
    & \alpha_{13}J_{11}J_{13}\dot{J}_{13}-\alpha_{13}J_{13}\dot{J}_{31}J_{33}+\alpha_{13}J_{13}J_{31}\dot{J}_{33} \bigr) ,
\end{align}
\begin{align}
\label{eq:J13_visc}
    \ddot{J}_{13,\mathrm{v}} = & -\frac{\hat{T}_0}{\Omega L^2 J^{\gamma+1}} \bigl(
    2\alpha_{11}\dot{J}_{13}J_{31}^2-2\alpha_{11}\dot{J}_{11}J_{31}J_{33}-\alpha_{13}J_{11}J_{13}\dot{J}_{11}+ \nonumber \\
    & \alpha_{13}J_{11}^2 \dot{J}_{13}+\alpha_{13}J_{11}\dot{J}_{31}J_{33}-\alpha_{13}J_{11}J_{31}\dot{J}_{33} \bigr) ,
\end{align}
\begin{align}
\label{eq:J21_visc}
    \ddot{J}_{21,\mathrm{v}} = & -\frac{\hat{T}_0}{\Omega L^2 J^{\gamma+1}} \bigl(
    -\alpha_{21}\dot{J}_{23}J_{31}J_{33}+\alpha_{21}\dot{J}_{21}J_{33}^2+ \nonumber \\ 
    & \alpha_{23}J_{13}^2\dot{J}_{21}-\alpha_{23}J_{11}J_{13}\dot{J}_{23} \bigr) ,
\end{align}
\begin{align}
\label{eq:J23_visc}
    \ddot{J}_{23,\mathrm{v}} = & -\frac{\hat{T}_0}{\Omega L^2 J^{\gamma+1}} \bigl(
    \alpha_{21}\dot{J}_{23}J_{31}^2-\alpha_{21}\dot{J}_{21}J_{31}J_{33}- \nonumber \\ 
    & \alpha_{23}J_{11}J_{13}\dot{J}_{21}+\alpha_{23}J_{11}^2\dot{J}_{23} \bigr) ,
\end{align}
\begin{align}
\label{eq:J31_visc}
    \ddot{J}_{31,\mathrm{v}} = & -\frac{\hat{T}_0}{\Omega L^2 J^{\gamma+1}} \bigl(
    -\alpha_{31}\dot{J}_{11}J_{13}J_{33}+\alpha_{31}J_{11}\dot{J}_{13}J_{33}+\alpha_{31}\dot{J}_{31}J_{33}^2-\nonumber \\
    & -\alpha_{31}J_{31}J_{33}\dot{J}_{33}+2\alpha_{33}J_{13}^2\dot{J}_{31}-2\alpha_{33}J_{11}J_{13}\dot{J}_{33} \bigr) ,
\end{align}
\begin{align}
\label{eq:J33_visc}
    \ddot{J}_{33,\mathrm{v}} = & -\frac{\hat{T}_0}{\Omega L^2 J^{\gamma+1}} \bigl(
    \alpha_{31}\dot{J}_{11}J_{13}J_{31}-\alpha_{31}J_{11}\dot{J}_{13}J_{31}-\alpha_{31}J_{31}\dot{J}_{31}J_{33}+\nonumber \\
    & \alpha_{31}J_{31}^2 \dot{J}_{33}-2\alpha_{33}J_{11}J_{13}\dot{J}_{31}+2\alpha_{33}J_{11}^2\dot{J}_{33} \bigr).
\end{align}
Neglecting any viscous spreading of the ring, which occurs over very long timescales, the usual background equilibrium ring is given by 
\begin{equation}
    \frac{\hat{T}_0}{J^\gamma L^2} = \nu^2 \epsilon \quad \text{and} \quad C_z = \frac{1}{2\Omega}\left(\kappa^2 J_{11}-\nu^2\epsilon J_{33}\right).
\end{equation}
We now linearise the viscous ring model equations about this approximate equilibrium and retain only the components which break the midplane symmetry and hence correspond to tilting modes. Taking the Keplerian resonant case and choosing units such that $\Omega=\kappa=\nu=1$, this yields
\begin{align}
\label{eq:linear_J13_visc}
    \ddot{J}_{13} = & 2 \dot{J}_{23}+2 S J_{13}-\epsilon J_{31}-\alpha_{13}\bigl(\epsilon\dot{J}_{31}+\dot{J}_{13} \bigr) , \\
\label{eq:linear_J31_visc}
    \ddot{J}_{31} = & -J_{31} - \epsilon J_{13}-\epsilon\alpha_{31}(\dot{J}_{13}+\epsilon\dot{J}_{31}) , \\
\label{eq:linear_J23_visc}
    \ddot{J}_{23} = & -2\dot{J}_{13}-\alpha_{23}\bigl(\frac{3}{2}J_{13}+\frac{\epsilon^2}{2}J_{13}+\dot{J}_{23}\bigr)-\alpha_{21}\bigl(\frac{3}{2}\epsilon J_{31}+\frac{\epsilon^3}{2}J_{31}\bigr).
\end{align}
Now we wish to capture the fast oscillation of the tilting modes and also the slow viscous modification of the amplitude and phase. As such, we exploit the multiple timescales $T=\epsilon t$ and look for asymptotic solutions with the form 
\begin{equation}
    J_{ij} = J_{ij,0}(t,T)+\epsilon J_{ij,1}(t,T)+\mathcal{O}(\epsilon^2).
\end{equation}
Now the temporal derivative becomes $d_t = \partial_t+\epsilon\partial_T$. Furthermore, for weak damping we want the effect of viscosity to manifest at order $\epsilon$. Therefore we take $\alpha_{ij} = \epsilon\alpha_{ij}^{\prime}$ with $\alpha_{ij}^\prime=\mathcal{O}(1)$. Inserting these ansatz into equations \eqref{eq:linear_J13_visc}--\eqref{eq:linear_J23_visc} and grouping in orders of the aspect-ratio gives a series of simpler equations. At leading order we find
\begin{align}
\label{eq:J_130_visc}
    & \ddot{J}_{13,0} = 2 \dot{J}_{23,0}+2S J_{13,0}, \\
\label{eq:J_310_visc}
    & \ddot{J}_{31,0} = -J_{31,0}, \\
\label{eq:J_230_visc}
    & \ddot{J}_{23,0} = -2\dot{J}_{13,0}.
\end{align}
Combining equations \eqref{eq:J_130_visc} and \eqref{eq:J_230_visc} to eliminate the $J_{23,0}$ term and taking the Keplerian shear rate $S=3/2$ yields
\begin{equation}
    \label{eq:J_130_visc_harmonic}
    \ddot{J}_{13,0} = -J_{13,0}
\end{equation}
Thus the solutions for the tilt and shear are simply two decoupled oscillators with unit frequency such that 
\begin{align}
    & J_{13,0} = \mathrm{Re}\left[A(T)e^{-it}\right], \\
    & J_{31,0} = \mathrm{Re}\left[B(T)e^{-it}\right], \\
    & J_{23,0} = \mathrm{Re}\left[C(T)-2iA(T)e^{-it}\right]
\end{align}
where $A$, $B$ and $C$ are complex amplitudes which evolve over the slow timescale. In order to determine the evolutionary equations for these amplitudes we must go to next order in $\epsilon$ where
\begin{align}
    & \ddot{J}_{13,1}-2\dot{J}_{23,1}-2SJ_{13,1} = -J_{31,0}-\alpha_{13}^\prime \dot{J}_{13,0}+2\partial_T \left(J_{23,0}-\dot{J}_{13,0}\right) , \\
    & \ddot{J}_{31,1}+J_{31,1} = -J_{13,0}-2\partial_T\dot{J}_{31,0} , \\
    & \ddot{J}_{23,1}+2\dot{J}_{13,1} = -2\partial_T \left( \dot{J}_{23,0}+ J_{13,0}\right)-\alpha_{23}^\prime(\frac{3}{2}J_{13,0}+\dot{J}_{23,0}).
\end{align}
Inserting the leading order results and combining equations yields
\begin{align}
    & \ddot{J}_{13,1}+J_{13,1} = [i(\alpha_{13}^\prime+\alpha_{23}^\prime)A-B+2i\partial_T A]e^{-it}+2\partial_T C , \\
    & \ddot{J}_{31,1}+J_{31,1} = [-A+2i\partial_T B]e^{-it}.
\end{align}
The left hand sides are the same as the leading order free oscillators given by equations \eqref{eq:J_310_visc} and \eqref{eq:J_130_visc_harmonic}. In order to avoid a resonant secular growth we must therefore eliminate forcing terms on the right hand side proportional to $e^{-it}$. This solvability condition imposes the desired evolutionary equations
\begin{align}
\label{eq:A_visc_evolution}
    & \partial_T A = -\frac{i}{2}B-\frac{(\alpha_{13}^\prime+\alpha_{23}^\prime)}{2}A, \\
\label{eq:B_visc_evolution}
    & \partial_T B = -\frac{i}{2}A.
\end{align}

%%%%%%%%%%%%%%%%%%%%%%%%%%%%%%%%%%%%%%%%%%%%%%%%%%

% Don't change these lines
\bsp	% typesetting comment
\label{lastpage}
\end{document}